\documentclass[useAMS,usenatbib]{mn2e}

\def\Zsol{\hbox{Z$_{\odot}$}}
\def\Msol{\hbox{M$_{\odot}$}}

\usepackage{epsfig,natbib,url,graphicx,subfig,float}
\usepackage{lscape,amsmath,amssymb}
\usepackage{booktabs,threeparttable}
\usepackage{longtable,deluxetable}
\usepackage{color}
\usepackage{soul} 
\usepackage{fancyvrb}
\usepackage{pifont}
\usepackage{ulem}
\usepackage{pdflscape}

\newcommand{\hii}{H~{\sc ii}}
\newcommand{\hei}{He~{\sc i}}
\newcommand{\heii}{He~{\sc ii}}

\newcommand{\eld}{$N_{\rm e}$}

\newcommand{\elt}{$T_{\rm e}$}

\newcommand{\op}{O$^+$}

\newcommand{\opp}{O$^{2+}$}

\newcommand{\np}{N$^+$}

\newcommand{\foiii}{[O~{\sc iii}]}

\newcommand{\foi}{[O~{\sc i}]}
\newcommand{\foii}{[O~{\sc ii}]}
\newcommand{\fsii}{[S~{\sc ii}]}
\newcommand{\fsiii}{[S~{\sc iii}]}

\newcommand{\fnii}{[N~{\sc ii}]}

\newcommand{\fneiii}{[Ne~{\sc iii}]}

\newcommand{\fariii}{[Ar~{\sc iii}]}
\newcommand{\ariv}{Ar~{\sc iv}}

\newcommand{\feiii}{Fe~{\sc iii}}

\newcommand{\hp}{H$^+$}

\newcommand{\ha}{H$\alpha$}
\newcommand{\hb}{H$\beta$}
\newcommand{\hg}{H$\gamma$}
\newcommand{\hd}{H$\delta$}

\title[Blue diffuse dwarf galaxies]{Blue diffuse dwarf galaxies: a clearer 
picture\thanks{The observations reported here were obtained at the MMT Observatory, a joint facility of the University of Arizona and the Smithsonian Institution.}}
\author[James et al. ]{Bethan L. James$^{1,2,3}$\thanks{E-mail:bjames@ast.cam.ac.uk}, Sergey E. Koposov$^{1}$, Daniel P. Stark$^{4}$, Vasily Belokurov$^{1}$,  
\newauthor Max Pettini$^{1}$, Edward W. Olszewski$^4$, \&\ Kristen B. W. McQuinn$^5$\\
$^{1}$Institute of Astronomy, University of Cambridge, Madingley Road, Cambridge, CB3 0HA\\
$^2$Cavendish Laboratory, University of Cambridge, 19 J.J. Thomson Avenue, Cambridge, CB3 0HE, UK\\
$^3$Kavli Institute for Cosmology, University of Cambridge, Madingley Road, Cambridge CB3 0HA, UK\\
$^{4}$Steward Observatory, The University of Arizona, 933 N Cherry Ave, Tucson, AZ, 85721, USA\\
$^5$University of Texas at Austin, McDonald Observatory, 2515 Speedway, Stop C1402, Austin, TX 78712, USA }
\begin{document}

\date{Accepted 2016 November 14th. Received in original form 2016 August 11th}

\pagerange{\pageref{firstpage}--\pageref{lastpage}} \pubyear{2014}

\maketitle

\label{firstpage}

\begin{abstract}
The search for chemically unevolved galaxies remains prevalent in the nearby Universe, mostly because these systems provide excellent proxies for exploring in detail the physics of high-$z$ systems. The most promising candidates are extremely metal-poor galaxies (XMPs), i.e., galaxies with $<1/10$ solar metallicity. However, due to the bright emission line based search criteria traditionally used to find XMPs, we may not be sampling the full XMP population. In 2014 we reoriented this search using only morphological properties and uncovered a population of $\sim$150 `blue diffuse dwarf (BDD) galaxies', and published a sub-sample of 12 BDD spectra. Here we present optical spectroscopic observations of a larger sample of 51 BDDs, along with their SDSS photometric properties. With our improved statistics, we use direct-method abundances to confirm that BDDs are chemically unevolved (7.43$<$12+log(O/H)$<$8.01), with $\sim$20\%\ of our sample classified as being XMP galaxies, and find they are actively forming stars at rates of $\sim$1--33$\times10^{-2}$~\Msol/yr in \hii\ regions randomly embedded in a blue, low-surface brightness continuum.  Stellar masses are calculated from population synthesis models and estimated to be in the range $\log(M_*/\Msol)\simeq$5--9. Unlike other low-metallicity star-forming galaxies, BDDs are in agreement with the mass-metallicity relation at low masses, suggesting they are not accreting large amounts of pristine gas relative to their stellar mass. BDD galaxies appear to be a population of actively star-forming dwarf irregular (dIrr) galaxies who fall within the class of low-surface brightness dIrr galaxies. Their ongoing star-formation and irregular morphology make them excellent analogues for galaxies in the early Universe.
\end{abstract}


\section{Introduction}
According to hierarchical formation, dwarf ($M_\star \lesssim10^9$ M$_\odot$) galaxies are the first systems to collapse and start forming stars, supplying the building blocks for the formation of more massive galaxies through merging and accretion. As remnants of this process, present-day dwarfs may have been sites of the earliest star-formation (SF) activity in the Universe. However, to date, all dwarf irregular (dIrr) and blue compact dwarf galaxies (BCDs) with star-formation histories studied from HST imaging of their resolved stellar population show stars older than 10~Gyr\citep[e.g.,][]{Aloisi:1999,Aloisi:2007,Izotov:2002,McQuinn:2010,Weisz:2011}.
Irrespective of such findings, these chemically unevolved systems, with young stellar populations superimposed upon older, underlying populations, are still used as nearby guides that provide essential information on early galaxy formation and stellar evolution which cannot be extracted from the faint systems now being detected at $z\sim$9--11 \citep[e.g.,][]{Ellis:2013,Pirzkal:2015,Oesch:2016}. 

Based on the mass-metallicity relation \citep[e.g.,][]{Berg:2012} and the galaxy luminosity function, we know that low mass, and hence low metallicity, objects should be abundant.  However, locating the most metal-poor galaxies (i.e., extremely metal poor with $<0.1$~\Zsol\ or 12+log(O/H)$<7.69$, `XMPs')\footnote{Within this paper, we refer to metallicity in terms of oxygen abundance, 12+log(O/H), and define solar metallicity, \Zsol, as 12+log(O/H)=8.69 \citep{Asplund:2009}.}  within this group is difficult and so far, only a small number have been detected \citep[e.g.,][]{Izotov:2012,Sanchez-Almeida:2016}. Their detection is hindered by their intrinsic faintness and the fact that previous magnitude-limited selection techniques relied on the presence of bright \hii\ regions, which will only be the case if the galaxies are undergoing a rapid period of star-formation, i.e., a starburst. The cause of this intense star-forming activity still remains unclear - while many of these high-luminosity, low-mass XMP galaxies show signs of interaction, suggesting that both their star-formation activity and metal content may be influenced by their environment, a systematic study of the environment surrounding 20 XMP galaxies by \citet{Filho:2015} found that only 25\%\ of their sample had undergone major mergers or interactions in the past.

In light of the low yield from spectroscopic searches, new methods for XMP detection have been explored - e.g., blind H{\sc i} surveys. It was via this specific method that AGC~198691, the most metal-deficient galaxy now known \citep[$Z\sim1/50$\,$Z_\odot$,][]{Hirschauer:2016} was discovered, along with another XMP galaxy, Leo~P  \citep[$Z\sim1/34$\,$Z_\odot$,][]{Skillman:2013,Giovanelli:2013}.  This latter XMP in particular has very low levels of star-formation \citep[i.e., $<1\times10^{-4}$~\Msol/yr][]{Rhode:2013,McQuinn:2015a} and an unconventional morphology, with H{\sc ii} regions embedded within very low surface brightness emission. Moreover, the low metal content of Leo~P is thought to be a result of inefficient star-formation and a loss of metals through outflows \citep{McQuinn:2015b} - i.e., a result of secular evolution rather than interaction.  Leo~P's discovery and diffuse appearance led us to speculate how common is this structure amongst XMPs and can we use its morphological properties to uncover more XMPs? 

This indeed turned out to be true and using this method, \citet{James:2015} (J15 hereafter) brought to light a new sample of dwarf galaxies: `Blue Diffuse Dwarf (BDD) Galaxies'. Using the morphological properties of Leo~P as a basis, a search was conducted on SDSS imaging data which uncovered $\sim$150 previously unstudied low surface-brightness star-forming galaxies .  These galaxies are faint, blue systems, each with isolated H{\sc ii} regions randomly scattered within a diffuse continuum.   The primary aim of this search was to find more XMPs like Leo~P and early optical long-slit spectroscopy of 12 candidates confirmed that $\sim$25\%\  of our sample were indeed XMPs.  Combined with SDSS imaging, our spectra reveal them to be nearby systems (5--120\,Mpc) with regions of recent and ongoing SF (i.e., ages of $\sim$7~Myr and star-formation rates of $\log(SFR/$M$_\odot$\,yr$^{-1}$)$\sim-3.5$ to $-0.5$) randomly distributed throughout their diffuse, chemically-unevolved gas.  As such, J15 concluded that BDDs may have properties that bridge the gap between quiescent dIrrs and starbursting BCDs, i.e., active star-formation in less structured systems. 

However, since J15 covered only a small fraction of the full sample, many of the conclusions regarding the BDD population as a whole suffered from small number statistics. We have now obtained long-slit observations for 51 out of 150 BDD galaxies and take this opportunity to re-visit the topic of BDDs within the context of other dwarf galaxies with a more statistically sound sample.  Dwarf galaxies appear to fall into groups/classifications depending on properties of star-formation, morphology, and also their chemical content. For example, BCDs are characterised by bright, compact starbursting regions, whereas dIrrs are quiescent, with multiple irregularly distributed \hii\, regions. XMP galaxies themselves also appear to fall into two groups - active, cometary-shaped systems and quiescent, diffuse structures. However, as we discuss in this paper, the classification of such systems may not be as clear cut and could instead be linked to the methodology used to detect them. 

The paper is structured as follows: in Section~\ref{sec:data} we present the sample of 51 BDD galaxies and describe observations and data reduction. Section~\ref{sec:results} presents the chemical (i.e., direct-method oxygen abundances and nitrogen-to-oxygen ratios), physical (i.e., luminosities, star-formation rates, ionisation parameters) and photometric (i.e., magnitudes, colours, stellar mass) properties of the galaxies. In Section~\ref{sec:disc} we discuss and explore these properties within the context of other dwarf galaxy samples via e.g., emission line diagnostics and the mass-metallicity relation, and in Section~\ref{sec:conc} we draw conclusions on our findings.
 \begin{figure*}
\includegraphics[scale=1]{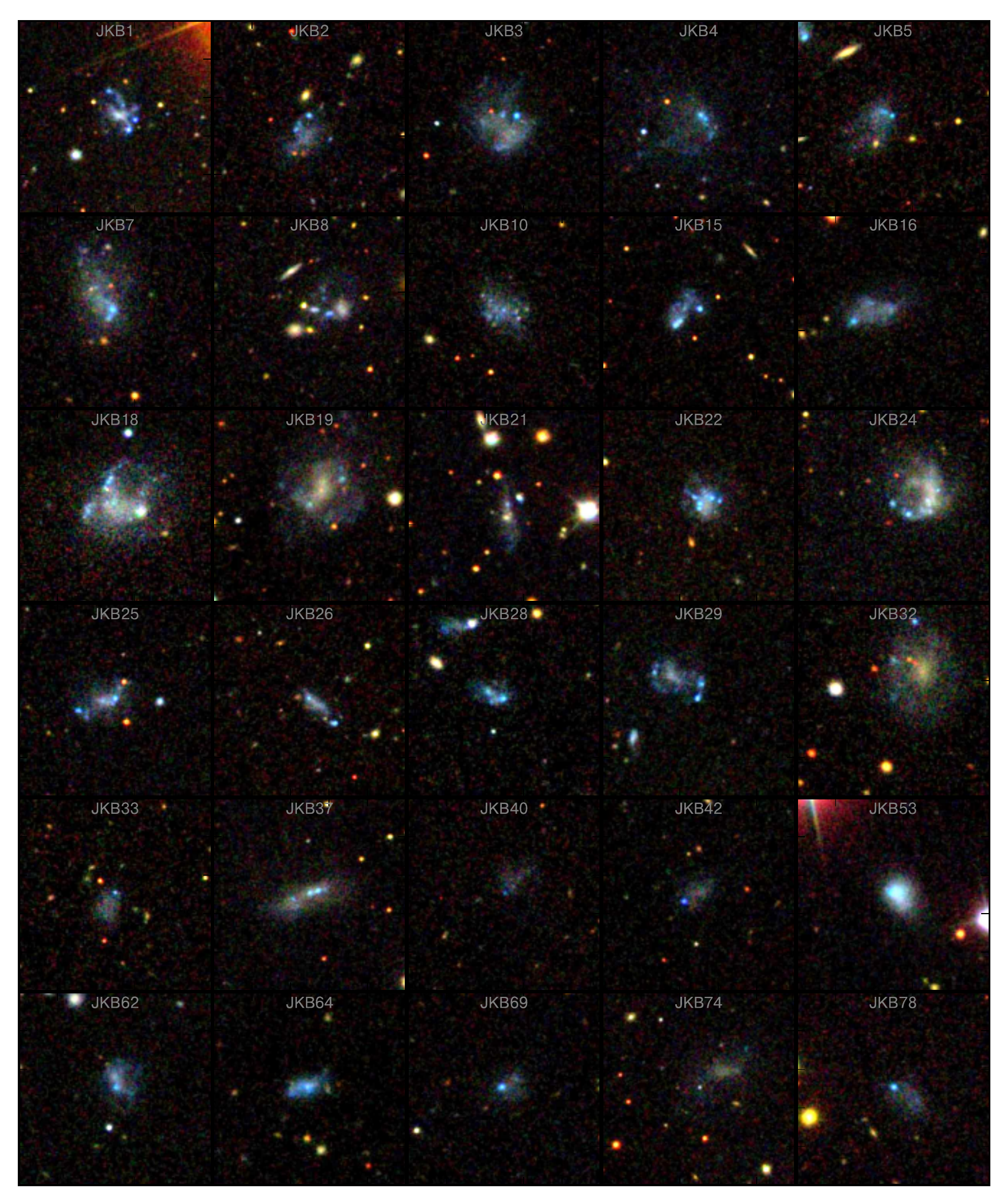}
\caption{SDSS images of the BDD galaxies presented in this paper. Information on each object can be found in Table~\ref{tab:gals}. Images were created from $gri$-band SDSS imaging and are $\sim$91\, acsec on each side. North is up and east is to the left.} 
\label{fig:gals}
\end{figure*} 
\begin{figure*}
\ContinuedFloat
\includegraphics[scale=1]{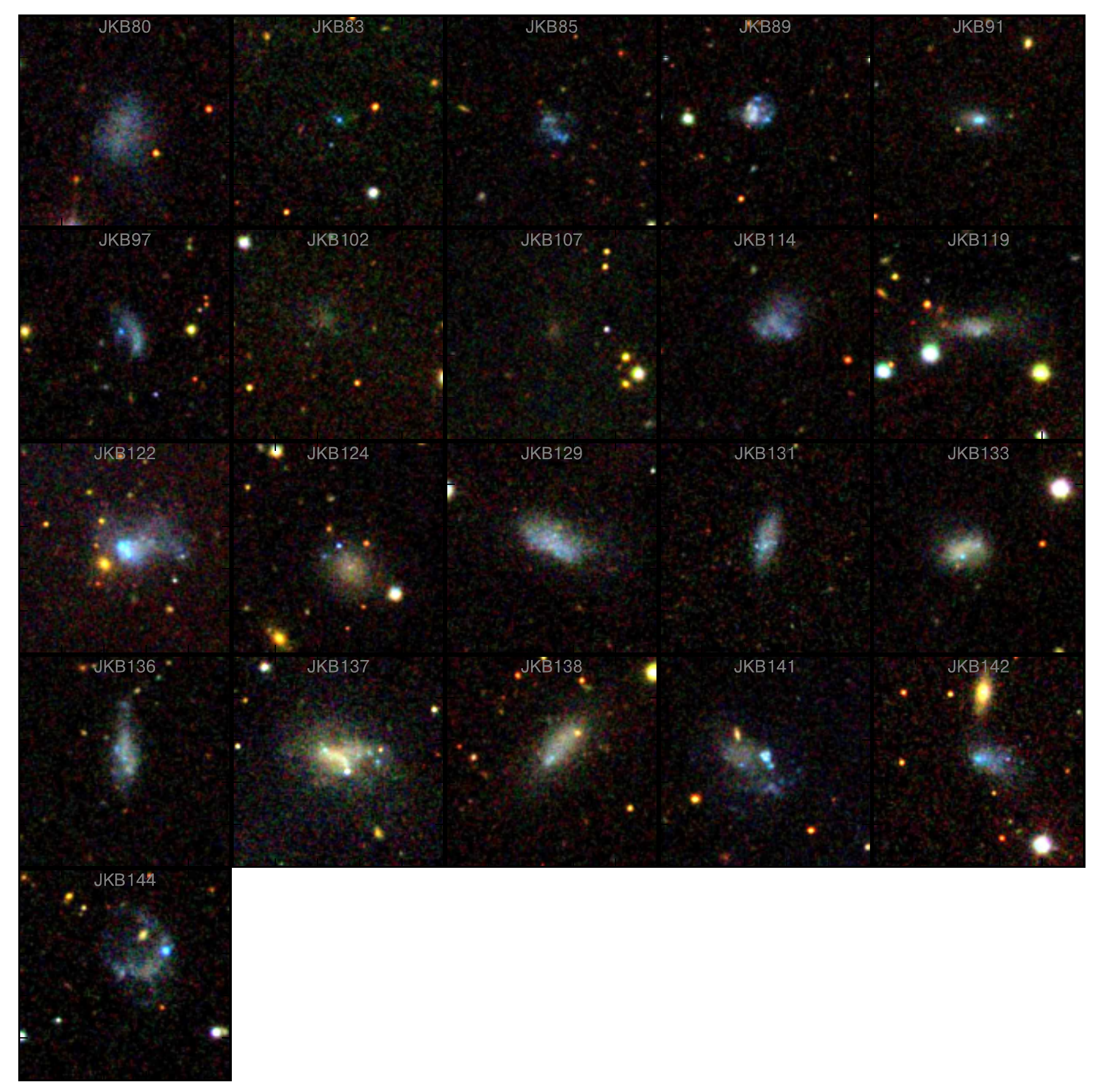}
\caption{-- continued.} 
\end{figure*}

\begin{table*}
\caption{Blue diffuse dwarfs galaxies in the current observed sample.  Identifications according to SDSS and AGC \citep[Arecibo General Catalog,][]{Haynes:2011} are also given where available.   Redshifts were derived from  fitting all the emission lines seen within the spectrum and converted into distances using the Hubble flow with respect to the CMB. Distances in parenthesis denote those which are uncertain due to peculiar velocities - see text for details. Distances noted with $^\star$ are for objects that are within 20 degrees of the Virgo cluster and will therefore have an additional uncertainty due to the Virgo flow.
SDSS images of each target can be found in Figure 1.  PA and $T_{\rm exp}$ refer to the position angle of the spectroscopic slit and the exposure time of the MMT observations, respectively. The JKB object number reflects the history of catalog creation.}
\begin{center}
\begin{tabular}{cccccccccc}
\hline
  Obj ID & SDSS ID& AGC ID & R.A. & Dec & $z$ & D (Mpc) & Slit PA ($\deg$) & $T_{\rm exp}$ (s) & UT Date of Obs.\\
 \hline
JKB~1 	&	J160753.65+090703.7	&	--	&	16:07:54	&	9:07:04	&	0.027860	&	122	&	135	&	2x600	&	28 Jan 2014	\\ 
JKB~2	&	J140406.04+310055.8	&	--	&	14:04:06	&	31:00:56	&	0.013988	&	60	&	42	&	3x900	&	28 Jan 2015	\\
JKB~3	&	J141708.66+134105.5	&	242011	&	14:17:09	&	13:41:06	&	0.016116	&	69	&	20	&	2x1200	&	28 Jan 2014	\\
JKB~4	&	J142532.35+523515.2	&	--	&	14:25:32	&	52:35:15	&	0.002019	&	(9)	&	133	&	3x900	&	28 Jan 2015	\\
JKB~5	&	 J085739.15+590256.4	&	--	&	8:57:39	&	59:02:56	&	0.003739	&	(16)	&	45	&	2x1200	&	28 Jan 2014	\\
JKB~7	&	J113453.11+110116.1	&	212838	&	11:34:53	&	11:01:16	&	0.002491	&	(11)	&	137	&	2x1200	&	28 Jan 2014	\\
JKB~8		& 	J083908.76+242130.4	&	749391	&	8:39:09	&	24:21:31	&	0.027810	&	122	& 110,67&	2x900, 2x1200	& 21, 22 Oct 2014 \\
JKB~10	& 	J001445.99+104846.9 	& 	--		& 	0:14:46	&	10:48:47	&	--		&	--	&	123	&	1x1200	&	21 Oct 2014 \\
JKB~15	& 	J000031.45+305209.3	& 	102729	& 	0:00:31	 &	30:52:09	&	0.015076	&	65	&	55 &	2x1200	&	21 Oct 2014	\\
JKB~16	&	J112710.87+391515.6	&	--	&	11:27:11	&	39:15:15	&	0.006797	&	29	&	74	&	3x800	& 24 Dec 2014	\\
JKB~18	&	J092127.17+072152.7	&	193816	&	9:21:27	&	7:21:53	&	0.004171	&	(18)	&	162	&	2x1200	&	28 Jan 2014	\\ 
JKB~19	&	J211309.36+043853.9	&	--	&	21:13:09	&	4:38:54	&	--	&	--	&	42	&	2x1200	&	22 Oct 2014	\\
JKB~21	&	J081246.92+043536.2	&	188949	&	8:12:47	&	4:35:36	&	0.033467	&	146	&	140	&	2x1200	&	22 Oct 2014	\\
JKB~22	&	 J114109.10+244505.2	&	749439	&	11:41:09	&	24:45:05	&	0.011464	&	49$^\star$	&	66	&	2x1200	&	24 Dec 2014	\\
JKB~24	&	J002211.19+111757.3	&	--	&	0:22:11	&	11:17:57	&	0.017668	&	76	&	51	&	2x1200	&	23 Dec 2014	\\
JKB~25	&	J010730.97-040129.0	&	--	&	1:07:31	&	-4:01:29	&	0.019589	&	85	&	175	&	3x1200	&	22 Oct 2014	\\
JKB~26	&	J141814.44+432249.7	&	--	&	14:18:14	&	43:22:50	&	0.012731	&	55	&	135	&	3x780	&	28 Jan 2015	\\
JKB~28	&	J005757.32-094119.2	&	--	&	0:57:57	&	-9:41:19	&	0.014728	&	63	&	30	&	2x1200	&	22 Oct 2014	\\
JKB~29 	&	J130110.39+365424.2	&	--	&	13:01:10	&	36:54:24	&	0.022526	&	98	&	105	&	1x947	&	28 Jan 2014	\\
JKB~32	&	J021111.29+281618.9	&	123128	&	2:11:11	&	28:16:19	&	--	&	--	&	175	&	2x1200	&	27 Jan 2014	\\
JKB~33	&	J005912.57+113007.8	&	--	&	0:59:13	&	11:30:08	&	--	&	--	&	60	&	1x1200	&	24 Dec 2014	\\
JKB~37	&	J062538.13+655228.0	&	--	&	6:25:38	&	65:52:28	&	0.013856	&	60	&	20	&	3x1200	&	27 Jan 2014	\\
JKB~40	&	J081459.45+451745.5	&	--	&	8:15:00	&	45:17:46	&	0.012695	&	55	&	85	&	2x1200	&	22 Oct 2014	\\
JKB~42	&	J092844.05+360100.3	&	--	&	9:28:44	&	36:01:00	&	0.030807	&	134	&	55	&	2x600	&	24 Dec 2014	\\
JKB~53	&	J115754.21+563816.7	&	--	&	11:57:54	&	56:38:17	&	0.001066	&	(5)	&	45	&	2x1200	&	28 Jan 2014	\\
JKB~62	&	J112331.57+571630.5	&	--	&	11:23:32	&	57:16:31	&	0.006574	&	28	&	95	&	2x1200	&	24 Dec 2014	\\
JKB~64	&	J133009.20+321715.9	&	--	&	13:30:09	&	32:17:16	&	0.002048	&	(9)	&	90	&	3x900	&	28 Jan 2015	\\
JKB~69	&	J100849.99+485351.1	&	--	&	10:08:50	&	48:53:51	&	0.014390	&	62	&	98	&	2x1200	&	24 Dec 2014	\\
JKB~74	& 	J081407.17+360906.7	&	--	&	8:14:07	&	36:09:07	&	--		&	--	&	96	&	1x1200	&	21 Oct 2014 \\
JKB~78	&	J092036.41+494031.3	&	--	&	9:20:36	&	49:40:31	&	0.001587	&	(7)	&	73	&	3x1200	&	28 Jan 2014	\\
JKB~80	&	 J092603.64+560915.5	&	--	&	9:26:04	&	56:09:15	&	0.002206	&	(11)	&	40	&	1x1200	&	28 Jan 2014	\\
JKB~83	&	J095549.64+691957.4	&	--	&	9:55:50	&	69:19:57	&	0.000185	&	(1)	&	92	&	2x1200	&	24 Dec 2014	\\
JKB~85	&	J095921.87+390856.4	&	--	&	9:59:22	&	39:08:56	&	0.022944	&	(10)	&	40	&	3x900	&	28 Jan 2015	\\
JKB~89	&	J101518.65+125934.1	&	205134	&	10:15:19	&	12:59:34	&	0.016585	&	72	&	10	&	3x720	&	24 Dec 2014	\\
JKB~91	&	J104701.47+273825.6	&	--	&	10:47:01	&	27:38:25	&	0.005110	&	22	&	90	&	3x720	&	24 Dec 2014	\\
JKB~97	&	J120920.02+242649.5	&	749452	&	12:09:20	&	24:26:50	&	0.008217	&	35$^\star$	&	135	&	2x800	&	24 Dec 2014	\\
JKB~102	&	J025139.17+334650.3	&	--	&	2:51:39	&	33:46:50	&	--	&	--	&	97	&	1x1200	&	27 Jan 2014	\\
JKB~107	&	J033410.64+064218.0	&	--	&	3:34:11	&	6:42:18	&	--	&	--	&	25	&	1x1200	&	27 Jan 2014	\\
JKB~114	&	J092423.79+402257.2	& 	--	&	9:24:24	&	40:22:57	& 	--		&	--	&	96	&	2x1200	&	21,22 Oct 2014 \\
JKB~119	&	J001722.78+174156.9	&	--	&	0:17:23	&	17:41:57	&	0.003161		&	(13)	&	159	&	2x1200	&	21 Oct 2014 \\
JKB~122	&	J122542.55+264834.5	&	749236	&	12:25:42	&	26:48:35	&	0.000371	&	(2)	&	146	&	2x1200	&	24 Dec 2014	\\
JKB~124	&	J014850.49+220641.6	&	--		&	1:48:50	&	22:06:42	&	--	&	--	&	150		&	1x1200	&	21 Oct 2014	\\
JKB~129	&	J002041.45+083701.2	&	103419	&	0:20:41	&	8:37:01	&	---	&	---	&	56	&	3x1200	&	24 Dec 2014	\\
JKB~131	&	J013813.59+105825.6	&	--	&	1:38:14	&	10:58:26	&	0.010041	&	43	&	55	&	2x1200	&	22 Oct 2014 \\
JKB~133	&	J023711.21+205238.5	&	--	&	2:37:11	&	20:52:39	&	0.013318	&	57	&	57	&	2x1200	&	21 Oct 2014\\
JKB~136	&	J020521.94-054153.2	&	--	&	2:05:22	&	-5:41:53	&	0.017456	&	75	&	40	&	2x1200	&	21 Oct 2014\\
JKB~137	&	J215817.93-025404.8	&	--	& 21:58:18	&	-2:54:05 	 & 0.016668 & 72	& 4 & 2x1200 &	22 Oct 2014 \\
JKB~138	&	J014610.53+284901.4	& 111473	&	1:46:11	&	28:49:02	&	 0.012198	& 52	& 45 & 2x1200 &	22 Oct 2014\\
JKB~141	&	J002004.10+083021.5	&	103622	&	0:20:04	&	8:30:22	&	0.018561	&	80	&	145	&	2x1200	&	24 Dec 2014	\\
JKB~142	&	J014548.23+162240.6	&	--	&	1:45:48	&	16:22:40	&	0.002304	&	(10)	&	115	&	2x1200	&	24 Dec 2014	\\
JKB~144	&	J233102.23+222644.3	&	332237	&	23:31:02	&	22:26:44	&	0.022897	&	99	&	48	&	2x1200	&	23Dec 2014	\\
  \hline
\end{tabular}
\end{center}
\label{tab:gals}
\end{table*}%

\section{Overview of Selection Criteria}
A full description of the selection criteria used to create this sample can be found in J15. Here we provide a brief overview of the criteria used to equip the reader with a better understanding of the sample's properties.  As mentioned previously, the selection criteria were primarily motivated by the morphology of Leo~P, a recently discovered XMP galaxy which, in SDSS images, appears as a very diffuse low surface brightness object with several embedded \hii\ regions. The main selection criteria adopted and applied to a local SQL (Structured Query Language) database of the SDSS DR9 catalog were the following:

\begin{itemize}
\item Presence of blue point sources  $-0.5<(g-r)_0<1$, $-2<(r-i)_0<0$ 
consistent with \hii\ regions. In order to avoid artefacts from bright stars, we required that the SDSS photometric flags for these objects would not have \textsc{SUBTRACTED} and \textsc{SATURATED} fields set\footnote{\url{https://www.sdss3.org/dr8/algorithms/flags\_detail.php}}. The objects having higher density of such sources within 10, 20, 30 arcseconds were ranked higher.
\item Presence of one or several of faint extended sources ($\rm modelmag_g>20$) within 10, 20, 30 arcseconds.
\item Absence of bright ($\rm modelmag_g<18$) galaxy nearby.
\item Absence of known LEDA\footnote{Lyon-Meudon Extragalactic Database: \url{http://leda.univ-lyon1.fr/}} source or bright UCAC\footnote{USNO CCD Astrograph Catalog: \url{http://www.usno.navy.mil/USNO/astrometry/optical-IR-prod/ucac}} star nearby. 
\item High galactic latitude.
\end{itemize}

After deselecting all known sources within NED and inspecting the list for promising cases, we obtained a final list of $\sim 150$ low surface brightness star-forming dwarf galaxies. Overall, with respect to the properties of Leo~P (our prototype for this search), the JKB objects are found to have similarly low luminosities and metallicities, but at the same time are more massive and with more active star-formation.  We discuss each of these properties in detail in Sections~\ref{sec:results} and \ref{sec:disc}.

\section{Observations And Data Processing}\label{sec:data}
At present, our full sample consists of $\sim$150 BDD galaxies,  of which we have obtained optical spectroscopy for 51 galaxies, as listed in Table~\ref{tab:gals}. The sample considered here includes both the 12 galaxies originally presented in J15 and 39 new BDD candidates.  The table is listed in order of their JKB number\footnote{Due to cross-identification purposes within NED and SIMBAD, it was necessary to replace the `KJ' acronym used in J15 with `JKB'. We stress to the reader that all objects presented here (which includes those presented in J15) will be listed within NED and SIMBAD as `JKB' \textit{not} `KJ', and will be referred to as `JKB objects' hereafter.}, along with general properties such as coordinates and redshift, and observational details such as the position angle of the spectrograph slit and exposure time.  The JKB objects are not numbered with respect to their RA - instead the JKB number reflects the history of catalog creation. Observations were made with the Multiple Mirror Telescope (MMT)  blue channel spectrograph on the dates indicated in Table~\ref{tab:gals}, using the $1''\times180''$ slit with medium resolution grating (300 grooves mm$^{-1}$). This gives a spatial scale along the slit of 0.6 arcsec pixel$^{-1}$, a spectral range of 3500--8000~\AA, and a spectral resolution of $\sim7$~\AA\ [full width at half maximum (FWHM)].

As in J15, the processing of the spectra was performed using a custom pipeline that follows all the standard reduction steps.  As such, and in order to maintain a homogeneous sample of data, all observations have been reduced together using the same pipeline, i.e., including those presented in J15. Raw data were bias subtracted and flat-fielded, and wavelength calibration was performed using the lamp spectra recorded immediately after or before each science exposure.  After subtracting the sky spectrum from each science frame, the spectra of science objects were extracted by summing the flux along the slit within an aperture.  Extracted spectra from multiple exposures were combined using weighted average together with $\sigma$-clipping to exclude residuals from cosmic rays. The flux calibration of the spectra was performed using spectrophotometric standard stars observed throughout each night with the instrument set-up described above. Uncertainties resulting from flux calibration were estimated to be $\sim$10\%\ and were incorporated into the error spectrum. We adopt this somewhat conservative error in order to account for the effects of systematic uncertainties, such as not fully photometric conditions, seeing variation throughout the night, possible slit losses etc. We note that in order to avoid large systematic slit losses due to differential refraction \citep{Fillippenko:1982}, all the objects below airmass of $\sim 1.1$ were observed at the parallactic angle (which is listed as the position angle in Table~\ref{tab:gals}). In the cases where multiple \hii\ regions were clearly visible in the two-dimensional spectra, their spectra were separately extracted. Such cases are denoted has JKB X.1, JKB X.2 etc., where X denotes the JKB catalogue number. 

The emission lines detected within each spectrum were fitted with a single Gaussian component (multiple Gaussian components were not necessary).  The resultant redshifts are given in Table~\ref{tab:gals} whereas the integrated line fluxes for each emission line are given in Table~\ref{tab:fluxes_all}. The distances have been derived assuming the Hubble flow ($H_0$=70~km\,s$^{-1}$\,Mpc$^{-1}$) with respect to the CMB \citep{Hinshaw:2009} and were found to range from $\sim$1--150~Mpc. Due to peculiar motions we expect that distances and distance dependent quantities will be unreliable for D$<$10--20\,Mpc \citep[see e.g.,][]{Tonry:2000}. For this paper we decided to focus on the sample of objects with with distances larger than 20\,Mpc, as for those objects the uncertainty (90\%\ interval) on the distance modulus is $\lesssim$1 magnitude, as tested using the sample of nearby galaxies from the HyperLEDA catalogue \citep{Makarov:2014}. This distance uncertainty for objects with D$\gtrsim$20~Mpc should provide an additional 1~dex scatter of $\lesssim$0.5\,mag on photometric quantities (i.e., absolute magnitudes and surface brightnesses) and $\lesssim$0.3~dex on parameters like star formation rate, stellar mass etc. In tables presented in Sections~\ref{sec:results} we mark all the distance dependent quantities for objects with D$<$20\,Mpc by `$^\star$', to signal the possibility of significant uncertainty and do not include such quantities in the plots or related discussions.

At this point of our analysis, objects for which no strong emission lines (e.g., \ha, \hb, or \foiii) were detected were removed from our spectroscopic analysis (namely JKB~10, 19, 32, 33, 74, 80, 102, 107, 114, 124, 129). They are, however, included in the photometric analysis described in Section~\ref{sec:phot}.  Any separately extracted spectra of different positions within the objects that showed no evidence of emission lines were also excluded from our spectroscopic analysis. Spectra of all objects, and their separately extracted \hii\ region spectra, for those with both detected and un-detected emission lines, are shown in Fig~\ref{fig:allspec}.

We stress here that the non-emission line JKB objects are still considered as an important part of our BDD sample. This is because we employ a photometric rather than emission-line based selection technique and as such the JKB objects are expected to represent galaxies at different evolutionary phases. For example, the non-emission line objects may represent BDD galaxies that are observed during quiescent periods between star-forming events. 

A full spectroscopic analysis was undertaken on 40/51 objects and, if we account for spectra extracted from individual \hii\ regions, this amounts to 57 individual spectra. As in J15, line fluxes and their uncertainties were obtained by sampling the likelihood function with a Markov chain Monte Carlo method, in order to simultaneously fit all of the observed lines within the spectrum, while requiring the redshifts and line widths to be the same for all the lines. If a line was found to be below a 3$\sigma$ detection, we measure the upper limit on the line flux using the error spectrum and the expected line profile. In some cases, underlying stellar absorption was visible, i.e., when the \ha\ equivalent width (EW) was less than 100~\AA.  In these cases, Balmer line fluxes were corrected accordingly.

\section{Results}~\label{sec:results}
We list both the observed and dereddened line fluxes in Table~\ref{tab:fluxes_all}.  Reddening values were estimated from the observed F(\ha)/F(\hb) and F(\hb )/F(\hg) ratios (combined in a 3:1 ratio, respectively), adopting the case B recombination ratios (appropriate for gas with T = 10 000~K and \eld = 100 cm$^{-3}$) and the Large Magellanic Cloud (LMC) extinction curve \citep{Fitzpatrick:1999}. In cases where \hg\ was not detected, only the F(\ha)/F(\hb) ratio was employed.

\subsection{Chemical properties}\label{sec:abund}
\begin{figure*}
\includegraphics[scale=0.6]{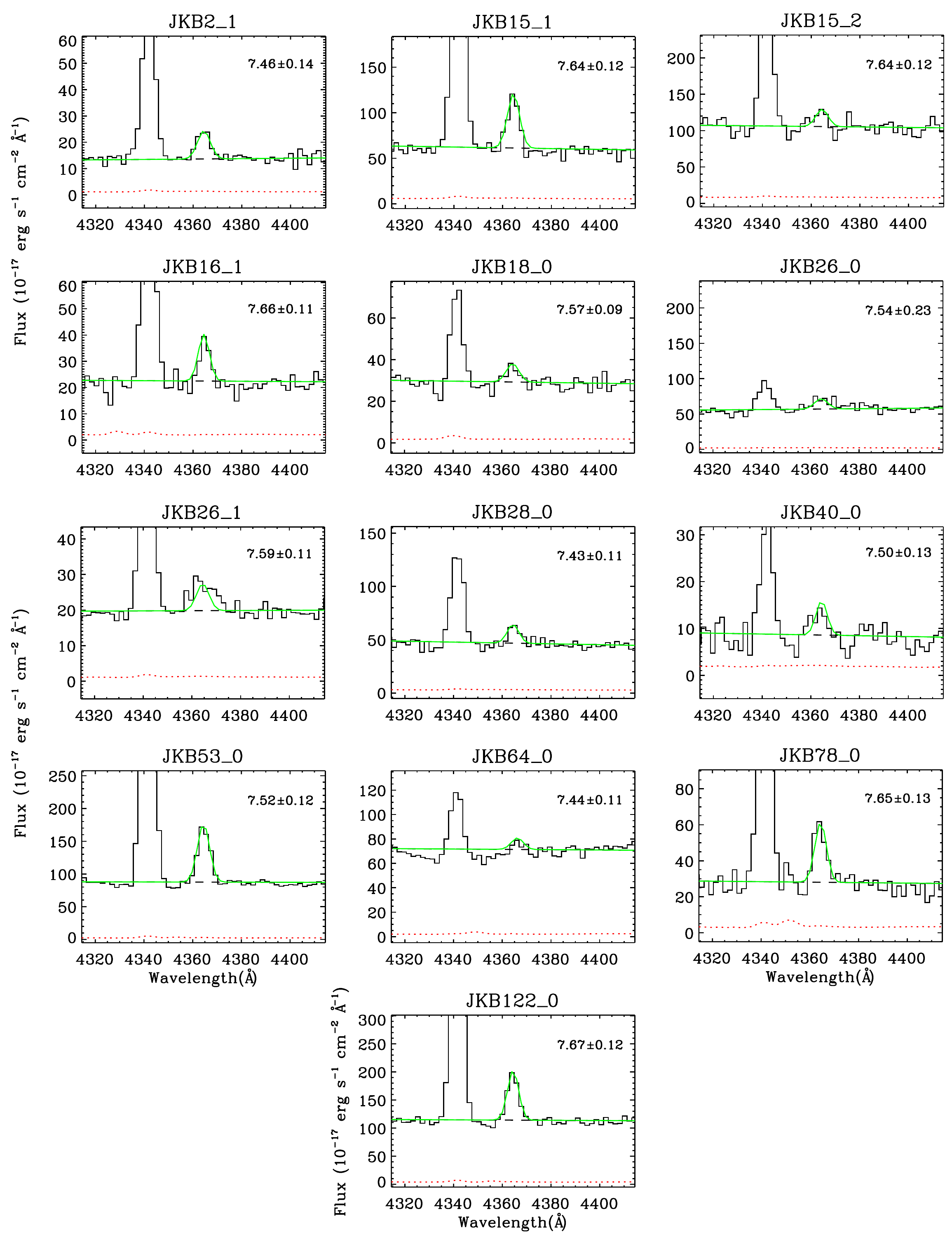}
\caption{The auroral line, \foiii~$\lambda$4363 profile for the confirmed extremely metal poor (i.e., 12+log(O/H)$<$7.69) BDDs in our sample. Each zoom-in shows the observed spectrum (black histogram), error (red-dashed), and fitted profile to the \foiii~$\lambda$4363 line (green). Measured fluxes are listed in Tables~\ref{tab:fluxes_all}, while the resultant chemical abundances can be found in Table~\ref{tab:abunds}. The 12+log(O/H) value of each object is inset within each panel.}
\label{fig:XMPs}
\end{figure*} 

Chemical abundances were calculated using the `direct method', where abundance measurements are based on the physical conditions of the gas (i.e., utilising electron temperature, \elt, and electron density, \eld) and extinction corrected line fluxes.  Our methodology differs slightly to that adopted by J15 in that we utilise the updated atomic data presented in \citet{Berg:2015} and assume that the ions are well-approximated by a 5-level atom.\footnote{We implement this via the IDL library \textsc{impro}: \url{https://github.com/moustakas/impro}} Each \hii\ region is modelled by three separate ionisation zones (low, medium, and high) and the abundance calculations for ions within each zone are made using the temperature within the respective zone. Here we deal with species from only the medium and low-ionisation zones. For the medium-ionisation zone, where \opp\ resides, we use \elt(\foiii) calculated from \foiii\, ($\lambda4959+\lambda5007$)/\foiii($\lambda4363$). \elt(\foiii) is derived using an iterative method alongside the calculation of electron densities \eld(\fsii) ($\lambda6716/\lambda6731$) and \eld(\foii) ($\lambda3729/\lambda3727$). 

Using this direct electron temperature, we then calculate \elt(\foii) for the low-ionisation zone, where \op\, resides. We calculate \elt(\foii) using the relationship between \elt(\foii) and \elt(\foiii) from \citet{Izotov:2006} derived from metal-poor emission line galaxies. While several similar relationships can be found in the literature \citep[e.g.,][]{Garnett:1992,Lopez-Sanchez:2012,Croxall:2016}, not all are well constrained within the high \elt\ (i.e., low metallicity) regime, where many of our objects lie. As discussed by \citet{Croxall:2016}, systematic offsets can result from \elt(\foii) relations that utilise different atomic data, although such effects on our \op/\hp\ abundance determination are negligible since the uncertainty in \elt(\foiii) is the dominant factor.  Uncertainties in both values of \elt\ correspond to uncertainties in the relevant extinction-corrected fluxes, and are then propagated into the errors in the abundance values.

 We compute both \op/\hp and \np/\hp\ abundances using the low ionisation zone \elt\ and \eld. Total oxygen abundances are calculated from the sum of \opp\ and \op, and N/O abundance ratios are derived simply from \np/\op\ (a valid assumption given that both ions reside in the same ionisation zone and hence no ionisation correction factors are needed).

Table~\ref{tab:abunds} lists the resultant \elt(\foiii), \elt(\foii), \op/\hp, \opp/\hp, 12+log(O/H) and log(N/O) for the sample.  We also list the weighted mean \eld\ calculated from \eld(\foii) and \eld(\fsii) and adopting \elt(\foiii). Here we choose to use \elt(\foiii) over \elt(\foii) because (i) \elt(\foiii) is more robustly (i.e., directly) measured than \elt(\foii) and (ii) \eld\ measurements made using \elt(\foii) rather than \elt(\foiii) were found to differ by $<10$~cm$^{-3}$ due to the weak dependence of \eld\ on \elt.

Direct calculations of the oxygen abundance (i.e., from detections of the \foiii~$\lambda$4363 line) were available for $\sim$50\%\ of the sample (i.e., 26/51 objects).  For those spectra where \foiii~$\lambda$4363 was undetected, we list the oxygen abundances as lower limits. Since the \foiii\ emission lines in question are coolants in the \hii\ regions, in the abundance range relevant here an upper limit on the flux ratio of \foiii~$\lambda 4363$/\foiii~$\lambda 5007$ (deduced from the noise level at the redshifted wavelength of $\lambda 4363$) translates into an upper limit on the temperature and thereby a lower limit on the oxygen abundance \citep{Stasinska:2012}.
For those spectra where \fnii~$\lambda$6584 was undetected, we list N/O ratios as upper limits (due to the upper limit on the flux detection). N/O abundance ratios are not listed for galaxies with lower limit oxygen abundances.  Out of those for which direct-method abundances were obtainable, oxygen abundances were found to be in the range  7.43$<$12+log(O/H)$<$8.01, with 11 out of 26 objects falling within the XMP classification (i.e., 0.1$<$\Zsol, 12+log(O/H)$<$7.69).  This accounts for approximately 20\%\ of the total sample presented here, which is in rough agreement with the fraction of XMP objects reported in J15 (25\%).  We show the \foiii~$\lambda$4363 profile for each object classified as an XMP in Figure~\ref{fig:XMPs}. Nitrogen-to-oxygen ratios were found to lie in the range $-1.88<$log(N/O)$<-1.39$. 


\begin{table*}
\begin{center}
\begin{footnotesize}
\begin{tabular}{|l|cccccccc|}
\hline
JKB ID & \eld(\foii,\fsii ) & \elt(\foiii)  & \elt(\foii) & \op/\hp & \opp /\hp  & 12+log(O/H) & log(N/O) & [O/H] \\
 & (cm$^{-3}$)  & (K) & (K) & ($\times10^{-5}$) & ($\times10^{-5}$) & & & \\ 
\hline
1.0& 
330$\pm$170 & 
$<$16100&
$<$14400&
$>$   3.66&
$>$   3.24&
$>$   7.84&
--&
$>$  -0.85\\
1.1& 
790$\pm$130 & 
15900$\pm$1870&
14300$\pm$743&
   1.41$\pm$   0.37&
   4.41$\pm$   1.58&
   7.76$\pm$   0.12&
$<$  -2.34&
  -0.93$\pm$   0.13\\
2.0& 
810$\pm$150 & 
$<$11600&
$<$11400&
$>$   3.44&
$>$   3.00&
$>$   7.81&
--&
$>$  -0.88\\
2.1& 
730$\pm$290 & 
17200$\pm$2380&
14700$\pm$573&
   0.48$\pm$   0.12&
   2.40$\pm$   0.94&
   7.46$\pm$   0.14&
$<$  -2.07&
  -1.23$\pm$   0.15\\
3.0& 
210$\pm$65 & 
13200$\pm$1820&
12800$\pm$1330&
   2.34$\pm$   1.17&
   8.00$\pm$   4.19&
   8.01$\pm$   0.18&
  -1.59$\pm$   0.14&
  -0.68$\pm$   0.19\\
3.1& 
550$\pm$130 & 
$<$18800&
$<$15000&
$>$   5.46&
$>$   1.27&
$>$   7.83&
--&
$>$  -0.86\\
4.0& 
160$\pm$47 & 
14100$\pm$2040&
13400$\pm$1260&
   2.67$\pm$   1.05&
   2.71$\pm$   1.26&
   7.73$\pm$   0.13&
  -1.59$\pm$   0.11&
  -0.96$\pm$   0.14\\
5.0& 
290$\pm$110 & 
$<$14000&
$<$13300&
$>$   3.04&
$>$   1.67&
$>$   7.67&
--&
$>$  -1.02\\
7.0& 
150$\pm$48 & 
13400$\pm$1440&
12900$\pm$1010&
   3.25$\pm$   1.13&
   4.93$\pm$   1.83&
   7.91$\pm$   0.11&
  -1.46$\pm$   0.10&
  -0.78$\pm$   0.12\\
8.11& 
46$\pm$27 & 
14800$\pm$1710&
13800$\pm$903&
   2.60$\pm$   0.78&
   4.06$\pm$   1.50&
   7.82$\pm$   0.11&
  -1.58$\pm$   0.10&
  -0.87$\pm$   0.12\\
8.12& 
1.1$^{+2.0}_{-1.1}$ & 
$<$28000&
$<$10200&
$>$  16.76&
$>$   0.77&
$>$   8.24&
--&
$>$  -0.45\\
8.2& 
490$\pm$370 & 
$<$24200&
$<$13400&
$>$   2.03&
$>$   0.87&
$>$   7.46&
--&
$>$  -1.23\\
8.3& 
180$\pm$120 & 
13300$\pm$1710&
12900$\pm$1220&
   2.19$\pm$   0.89&
   5.40$\pm$   2.34&
   7.88$\pm$   0.14&
  -1.45$\pm$   0.12&
  -0.81$\pm$   0.15\\
15.1& 
400$\pm$120 & 
17100$\pm$2190&
14700$\pm$560&
   1.15$\pm$   0.28&
   3.26$\pm$   1.20&
   7.64$\pm$   0.12&
  -1.70$\pm$   0.13&
  -1.05$\pm$   0.13\\
15.2& 
570$\pm$130 & 
16800$\pm$3010&
14600$\pm$858&
   2.40$\pm$   0.68&
   1.95$\pm$   0.99&
   7.64$\pm$   0.12&
  -1.78$\pm$   0.14&
  -1.05$\pm$   0.13\\
16.1& 
1.7$^{+6.8}_{-1.7}$ & 
15800$\pm$1890&
14300$\pm$784&
   1.51$\pm$   0.41&
   3.02$\pm$   1.11&
   7.66$\pm$   0.11&
  -1.47$\pm$   0.09&
  -1.03$\pm$   0.12\\
16.2& 
470$^{+640}_{-470}$ & 
$<$23600&
$<$13800&
$>$   4.70&
$>$   0.81&
$>$   7.74&
--&
$>$  -0.95\\
16.3& 
3.4$^{+18}_{-3.4}$ & 
$<$28000&
$<$10200&
$>$  14.73&
$>$   0.34&
$>$   8.18&
--&
$>$  -0.51\\
18.0& 
370$\pm$110 & 
19600$\pm$2860&
14900$\pm$164&
   2.07$\pm$   0.44&
   1.63$\pm$   0.62&
   7.57$\pm$   0.09&
  -1.67$\pm$   0.09&
  -1.12$\pm$   0.10\\
18.1& 
4.7$^{+25}_{-4.7}$ & 
$<$28000&
$<$10200&
$>$  37.76&
$>$   0.54&
$>$   8.58&
--&
$>$  -0.11\\
21.0& 
40$^{+46}_{-40}$ & 
$<$13000&
$<$12600&
$>$   5.76&
$>$   6.44&
$>$   8.09&
--&
$>$  -0.60\\
22.0& 
6.7$^{+6.8}_{-6.7}$ & 
13400$\pm$1240&
13000$\pm$869&
   2.11$\pm$   0.66&
   7.35$\pm$   2.44&
   7.98$\pm$   0.12&
  -1.41$\pm$   0.09&
  -0.71$\pm$   0.13\\
22.1& 
140$\pm$45 & 
14100$\pm$1500&
13400$\pm$934&
   2.04$\pm$   0.65&
   5.77$\pm$   2.07&
   7.89$\pm$   0.12&
  -1.60$\pm$   0.10&
  -0.80$\pm$   0.13\\
24.0& 
48$\pm$38 & 
$<$8970&
$<$8630&
$>$  28.20&
$>$  22.12&
$>$   8.70&
--&
$>$   0.01\\
24.1& 
29$^{+32}_{-29}$ & 
13500$\pm$1620&
13000$\pm$1120&
   3.63$\pm$   1.36&
   5.17$\pm$   2.09&
   7.94$\pm$   0.12&
  -1.57$\pm$   0.11&
  -0.75$\pm$   0.13\\
25.0& 
420$\pm$33 & 
13600$\pm$1370&
13100$\pm$931&
   2.11$\pm$   0.69&
   6.63$\pm$   2.32&
   7.94$\pm$   0.12&
  -1.57$\pm$   0.09&
  -0.75$\pm$   0.13\\
26.0& 
640$\pm$270 & 
25800$\pm$2530&
12300$\pm$2040&
   2.40$\pm$   1.81&
   1.08$\pm$   0.29&
   7.54$\pm$   0.23&
$>$  -2.12&
  -1.15$\pm$   0.23\\
26.1& 
320$\pm$77 & 
15900$\pm$2140&
14300$\pm$856&
   1.60$\pm$   0.45&
   2.30$\pm$   0.92&
   7.59$\pm$   0.11&
  -1.86$\pm$   0.13&
  -1.10$\pm$   0.13\\
28.0& 
740$\pm$190 & 
19100$\pm$2890&
15000$\pm$25.1&
   1.04$\pm$   0.22&
   1.65$\pm$   0.66&
   7.43$\pm$   0.11&
$<$  -2.26&
  -1.26$\pm$   0.12\\
29.0& 
360$\pm$140 & 
$<$13800&
$<$13200&
$>$   2.74&
$>$   2.03&
$>$   7.68&
--&
$>$  -1.01\\
29.2& 
840$^{+930}_{-840}$ & 
$<$12500&
$<$12300&
$>$   3.62&
$>$   3.09&
$>$   7.83&
--&
$>$  -0.86\\
37.0& 
130$\pm$45 & 
15200$\pm$1790&
14000$\pm$861&
   2.44$\pm$   0.72&
   4.33$\pm$   1.60&
   7.83$\pm$   0.11&
  -1.47$\pm$   0.09&
  -0.86$\pm$   0.12\\
40.0& 
180$\pm$120 & 
23300$\pm$2860&
13900$\pm$1440&
   2.06$\pm$   0.88&
   1.12$\pm$   0.35&
   7.50$\pm$   0.13&
  -1.48$\pm$   0.16&
  -1.19$\pm$   0.14\\
42.0& 
650$\pm$190 & 
16100$\pm$1800&
14400$\pm$681&
   0.71$\pm$   0.18&
   6.58$\pm$   2.27&
   7.86$\pm$   0.14&
  -1.54$\pm$   0.13&
  -0.83$\pm$   0.14\\
53.0& 
330$\pm$42 & 
17400$\pm$2250&
14800$\pm$476&
   0.79$\pm$   0.18&
   2.55$\pm$   0.94&
   7.52$\pm$   0.12&
  -1.61$\pm$   0.08&
  -1.17$\pm$   0.13\\
62.0& 
680$\pm$390 & 
$<$12800&
$<$12500&
$>$   2.83&
$>$   5.19&
$>$   7.90&
--&
$>$  -0.79\\
64.0& 
650$\pm$170 & 
22200$\pm$2780&
14400$\pm$1050&
   1.91$\pm$   0.61&
   0.83$\pm$   0.26&
   7.44$\pm$   0.11&
  -1.88$\pm$   0.15&
  -1.25$\pm$   0.12\\
69.0& 
380$\pm$110 & 
16000$\pm$2110&
14400$\pm$813&
   1.34$\pm$   0.38&
   4.54$\pm$   1.79&
   7.77$\pm$   0.13&
  -1.40$\pm$   0.11&
  -0.92$\pm$   0.14\\
78.0& 
430$\pm$110 & 
15500$\pm$1840&
14200$\pm$818&
   0.99$\pm$   0.28&
   3.53$\pm$   1.29&
   7.65$\pm$   0.13&
  -1.51$\pm$   0.10&
  -1.04$\pm$   0.14\\
83.0& 
220$\pm$86 & 
$<$13500&
$<$13000&
$>$   3.82&
$>$   2.02&
$>$   7.77&
--&
$>$  -0.92\\
85.0& 
92$\pm$60 & 
$<$12000&
$<$11800&
$>$   4.32&
$>$   5.85&
$>$   8.01&
--&
$>$  -0.68\\
89.0& 
290$\pm$74 & 
15100$\pm$2030&
14000$\pm$993&
   2.72$\pm$   0.86&
   4.20$\pm$   1.73&
   7.84$\pm$   0.12&
$<$  -2.29&
  -0.85$\pm$   0.13\\
91.0& 
130$\pm$21 & 
13000$\pm$1240&
12700$\pm$928&
   3.96$\pm$   1.33&
   5.85$\pm$   2.01&
   7.99$\pm$   0.11&
  -1.40$\pm$   0.10&
  -0.70$\pm$   0.12\\
97.0& 
35$\pm$29 & 
14800$\pm$2270&
13800$\pm$1200&
   2.08$\pm$   0.76&
   3.82$\pm$   1.82&
   7.77$\pm$   0.14&
  -1.58$\pm$   0.14&
  -0.92$\pm$   0.15\\
97.1& 
3500$^{+5400}_{-3500}$ & 
$<$28000&
$<$10200&
$>$  19.36&
$>$   0.50&
$>$   8.30&
--&
$>$  -0.39\\
119.0& 
1300$^{+1600}_{-1300}$ & 
$<$28000&
$<$10200&
$>$  34.30&
$>$   0.27&
$>$   8.54&
--&
$>$  -0.15\\
122.0& 
150$\pm$27 & 
15200$\pm$1750&
14000$\pm$843&
   1.31$\pm$   0.37&
   3.37$\pm$   1.22&
   7.67$\pm$   0.12&
  -1.56$\pm$   0.08&
  -1.02$\pm$   0.13\\
131.0& 
340$\pm$99 & 
$<$14100&
$<$13400&
$>$   4.64&
$>$   3.03&
$>$   7.88&
--&
$>$  -0.81\\
133.0& 
290$\pm$140 & 
$<$13500&
$<$13000&
$>$   5.24&
$>$   4.54&
$>$   7.99&
--&
$>$  -0.70\\
133.1& 
80$\pm$42 & 
14000$\pm$2030&
13400$\pm$1270&
   2.88$\pm$   1.16&
   4.99$\pm$   2.33&
   7.90$\pm$   0.14&
  -1.43$\pm$   0.12&
  -0.79$\pm$   0.15\\
136.0& 
720$\pm$260 & 
$<$12700&
$<$12400&
$>$   2.90&
$>$   4.94&
$>$   7.89&
--&
$>$  -0.80\\
137.0& 
15$^{+42}_{-15}$ & 
$<$27700&
$<$10600&
$>$  16.21&
$>$   0.87&
$>$   8.23&
--&
$>$  -0.46\\
138.2& 
840$\pm$380 & 
$<$23900&
$<$13600&
$>$   9.46&
$>$   0.81&
$>$   8.01&
--&
$>$  -0.68\\
141.0& 
200$\pm$47 & 
17500$\pm$2410&
14800$\pm$502&
   2.66$\pm$   0.62&
   2.51$\pm$   0.98&
   7.71$\pm$   0.10&
  -1.55$\pm$   0.08&
  -0.98$\pm$   0.11\\
142.0& 
84$\pm$42 & 
$<$10500&
$<$10400&
$>$   5.36&
$>$   5.89&
$>$   8.05&
--&
$>$  -0.64\\
144.0& 
600$\pm$43 & 
14000$\pm$1530&
13300$\pm$961&
   2.20$\pm$   0.71&
   6.88$\pm$   2.51&
   7.96$\pm$   0.12&
  -1.53$\pm$   0.09&
  -0.73$\pm$   0.13\\
144.1& 
390$\pm$370 & 
$<$28000&
$<$10200&
$>$  23.73&
$>$   1.28&
$>$   8.40&
--&
$>$  -0.29\\
\hline
\end{tabular}
\label{tab:abunds}
\end{footnotesize}
\end{center}

\caption{Ionic and elemental abundances for the BDD galaxies presented in this paper, derived from the emission line measurements given in
Table~\ref{tab:fluxes_all}. Electron density and temperatures are also listed.}\label{tab:abunds}
\end{table*}

\subsection{Physical properties}

\subsubsection{Star-formation rates and $L$(\ha)}
In Table~\ref{tab:SFR} we list the \ha\ luminosity and star-formation rates (SFRs) for our sample of BDD galaxies. For the cases with multiple \hii\ regions, we also list the total luminosities and SFRs.  Luminosities are calculated using the extinction-corrected \ha\ line fluxes (Table~\ref{tab:fluxes_all}) and the distances given in Table~\ref{tab:gals}. 
For objects with $D>20$~Mpc, total luminosities are in the range of 38.7$<$log($L_{\rm H\alpha}$/erg s$^{-1}$) $<$40.6, which lie in the intermediate range between quiescent dIrrs \citep{VanZee:2000} and starbursting BCDs \citep{GildePaz:2003}.  \ha\ luminosities are converted into SFRs using the \citet{Kennicutt:1998} (hereafter K98) prescription, except for low-luminosity cases, i.e., when $L$(\ha )$<2.5\times10^{39}$~erg\,s$^{-1}$, where we follow the prescription detailed in \citet{Lee:2009}. This latter prescription is essentially a recalibration the K98 relation to account for the under-prediction of SFRs by \ha\ compared to those from FUV fluxes within the low luminosity regime, under the assumption that the FUV traces the SFR in dwarf galaxies more robustly.  This recalibration is especially relevant for our sample as we lack FUV observations and are almost exclusively within the low SFR regime.
Total SFRs are seen to be in the range 0.1--3.3$\times10^{-1}$~\Msol/yr with a median of $3.5\times10^{-2}$~\Msol/yr (again excluding objects with $D<20$~Mpc).  SFRs for individual \hii\ regions have a slightly lower median of $2.4\times10^{-2}$~\Msol/yr.
These values are somewhat higher than those quoted within J15 due to the fact that in J15 we did not
use the \citet{Lee:2009} prescription and thus underestimated the SFRs for the galaxies with the lowest luminosities. Uncertainties in SFR values due to distance uncertainties are expected to be $\lesssim0.3$~dex. We discuss the star-formation properties of the JKB objects in relation to other dwarf galaxy types in Section~\ref{sec:disc}.
\begin{table*}
\caption{Physical properties of the new, extended sample of BDD galaxies: \ha\, luminosity and star-formation rate (listed for both individual \hii\ regions and for the entire galaxy), along with the age of the current star-forming population determined from the EW of \ha\ and \hb. Objects noted with `$^\star$' are those with uncertain distances, i.e., $D<20$~Mpc.}
\begin{center}
\begin{footnotesize}
\begin{tabular}{|l|ccccc|}
\hline
ID & $L$(\ha)  & SFR(\ha) & $L$(\ha)$_{total}$  & SFR(\ha)$_{total}$&  Age($EW$(\ha,\hb)) \\
   & ($\times10^{38}$ erg\,s$^{-1}$) & ($\times10^{-2}$ \Msol \,yr$^{-1}$)& ($\times10^{38}$ erg\,s$^{-1}$) & ($\times10^{-2}$ \Msol \,yr$^{-1}$) & (Myr)\\
\hline
JKB~  1.0&   31.48 $\pm$   3.58 &    2.49 $\pm$   0.28 &  124.98 $\pm$  10.12 &    9.88 $\pm$   0.80     & $<$11.25 \\  
JKB~  1.1 &   93.50 $\pm$   9.47 &    7.39 $\pm$   0.75 &                      &                         &  6.36 $\pm$  0.88 \\  
JKB~  2.0 &   15.09 $\pm$   1.65 &    2.17 $\pm$   0.39 &   44.36 $\pm$   3.91 &    4.48 $\pm$   0.48    & $<$ 8.51 \\  
JKB~  2.1 &   29.27 $\pm$   3.55 &    2.31 $\pm$   0.28 &                      &                         &  8.33 $\pm$  0.79 \\  
JKB~  3.0 &   56.25 $\pm$   8.91 &    4.44 $\pm$   0.70 &   70.05 $\pm$   9.27 &    6.50 $\pm$   0.79    &  4.27 $\pm$  0.48 \\  
JKB~  3.1 &   13.80 $\pm$   2.56 &    2.06 $\pm$   0.36 &                      &                         & $<$11.20 \\  
JKB~  4.0$^\star$ & 0.71 $\pm$   0.08 &    0.33 $\pm$   0.08 &    0.71 $\pm$   0.08 &    0.33 $\pm$   0.08   &  8.17 $\pm$  0.72 \\  
JKB~  5.0$^\star$&    1.09 $\pm$   0.12 &    0.43 $\pm$   0.10 &    1.09 $\pm$   0.12 &    0.43 $\pm$   0.10    & $<$ 9.82 \\  
JKB~  7.0$^\star$&    0.99 $\pm$   0.10 &    0.40 $\pm$   0.09 &    0.99 $\pm$   0.10 &    0.40 $\pm$   0.09    &  6.81 $\pm$  1.00 \\  
JKB~  8.1 &  109.12 $\pm$  11.04 &    8.62 $\pm$   0.87 &  258.99 $\pm$  18.05 &   22.33 $\pm$   1.51    &  7.18 $\pm$  0.94 \\  
JKB~  8.1 &    8.79 $\pm$   1.00 &    1.56 $\pm$   0.27 &                      &                         & $<$10.10 \\  
JKB~  8.2 &   18.89 $\pm$   1.99 &    2.50 $\pm$   0.46 &                      &                         & $<$11.46 \\  
JKB~  8.3 &  122.19 $\pm$  14.10 &    9.65 $\pm$   1.11 &                      &                         &  5.96 $\pm$  1.06 \\  
JKB~ 15.1 &   75.44 $\pm$   7.76 &    5.96 $\pm$   0.61 &  149.94 $\pm$  11.28 &   11.85 $\pm$   0.89    &  6.54 $\pm$  0.72 \\  
JKB~ 15.2 &   74.50 $\pm$   8.18 &    5.89 $\pm$   0.65 &                      &                         &  8.74 $\pm$  0.52 \\  
JKB~ 16.1 &    4.35 $\pm$   0.44 &    1.01 $\pm$   0.20 &    5.44 $\pm$   0.45 &    1.56 $\pm$   0.22    &  7.60 $\pm$  1.27 \\  
JKB~ 16.2 &    0.71 $\pm$   0.08 &    0.33 $\pm$   0.08 &                      &                         & $<$12.95 \\  
JKB~ 16.3 &    0.38 $\pm$   0.04 &    0.22 $\pm$   0.05 &                      &                         & $<$10.81 \\  
JKB~ 18.0$^\star$&    1.18 $\pm$   0.13 &    0.45 $\pm$   0.10 &    1.78 $\pm$   0.18 &    0.74 $\pm$   0.12    &  9.90 $\pm$  0.65 \\  
JKB~ 18.1$^\star$&    0.60 $\pm$   0.12 &    0.29 $\pm$   0.07 &                      &                         & $<$10.71 \\  
JKB~ 21.0 &   31.77 $\pm$   3.60 &    2.51 $\pm$   0.28 &   31.77 $\pm$   3.60 &    2.51 $\pm$   0.28    & $<$ 8.47 \\  
JKB~ 22.0 &   32.19 $\pm$   3.40 &    2.54 $\pm$   0.27 &   52.22 $\pm$   4.01 &    5.13 $\pm$   0.55    &  5.01 $\pm$  0.78 \\  
JKB~ 22.1 &   20.03 $\pm$   2.12 &    2.59 $\pm$   0.48 &                      &                         &  6.16 $\pm$  1.04 \\  
JKB~ 24.0 &  167.10 $\pm$  25.45 &   13.20 $\pm$   2.01 &  248.98 $\pm$  27.18 &   19.67 $\pm$   2.15    & $<$ 5.64 \\  
JKB~ 24.1 &   81.88 $\pm$   9.53 &    6.47 $\pm$   0.75 &                      &                         &  6.12 $\pm$  0.85 \\  
JKB~ 25.0 &   67.62 $\pm$   8.66 &    5.34 $\pm$   0.68 &   67.62 $\pm$   8.66 &    5.34 $\pm$   0.68    &  4.78 $\pm$  0.65 \\  
JKB~ 26.0 &    6.39 $\pm$   0.64 &    1.28 $\pm$   0.24 &   17.50 $\pm$   1.30 &    3.08 $\pm$   0.39    & 11.47 $\pm$  0.78 \\  
JKB~ 26.1 &   11.11 $\pm$   1.13 &    1.80 $\pm$   0.31 &                      &                         &  8.60 $\pm$  0.71 \\  
JKB~ 28.0 &   15.01 $\pm$   1.55 &    2.17 $\pm$   0.39 &   15.01 $\pm$   1.55 &    2.17 $\pm$   0.39    &  9.57 $\pm$  0.61 \\  
JKB~ 29.0 &   19.92 $\pm$   2.07 &    2.58 $\pm$   0.48 &   39.73 $\pm$   2.91 &    5.15 $\pm$   0.68    & $<$ 9.61 \\  
JKB~ 29.2 &   19.81 $\pm$   2.04 &    2.57 $\pm$   0.48 &                      &                         & $<$ 8.88 \\  
JKB~ 37.0 &   28.83 $\pm$   3.62 &    2.28 $\pm$   0.29 &   28.83 $\pm$   3.62 &    2.28 $\pm$   0.29    &  7.70 $\pm$  1.06 \\  
JKB~ 40.0 &    5.29 $\pm$   0.56 &    1.14 $\pm$   0.22 &    5.29 $\pm$   0.56 &    1.14 $\pm$   0.22    &  9.99 $\pm$  0.99 \\  
JKB~ 42.0 &  226.73 $\pm$  23.06 &   17.91 $\pm$   1.82 &  226.73 $\pm$  23.06 &   17.91 $\pm$   1.82    &  5.51 $\pm$  0.80 \\  
JKB~ 53.0$^\star$&    0.47 $\pm$   0.05 &    0.25 $\pm$   0.06 &    0.47 $\pm$   0.05 &    0.25 $\pm$   0.06    &  7.02 $\pm$  1.06 \\  
JKB~ 62.0 &    4.94 $\pm$   0.66 &    1.09 $\pm$   0.21 &    4.94 $\pm$   0.66 &    1.09 $\pm$   0.21    & $<$ 9.28 \\  
JKB~ 64.0$^\star$&    0.54 $\pm$   0.07 &    0.28 $\pm$   0.07 &    0.54 $\pm$   0.07 &    0.28 $\pm$   0.07    & 11.05 $\pm$  0.49 \\  
JKB~ 69.0 &   40.09 $\pm$   4.05 &    3.17 $\pm$   0.32 &   40.09 $\pm$   4.05 &    3.17 $\pm$   0.32    &  7.73 $\pm$  1.11 \\  
JKB~ 78.0$^\star$&    0.41 $\pm$   0.04 &    0.23 $\pm$   0.06 &    0.41 $\pm$   0.04 &    0.23 $\pm$   0.06    &  6.52 $\pm$  1.18 \\  
JKB~ 83.0$^\star$&    0.01 $\pm$   0.00 &    0.02 $\pm$   0.01 &    0.01 $\pm$   0.00 &    0.02 $\pm$   0.01    & $<$ 7.69 \\  
JKB~ 85.0$^\star$&   37.02 $\pm$   4.20 &    2.92 $\pm$   0.33 &   37.02 $\pm$   4.20 &    2.92 $\pm$   0.33    & $<$ 8.70 \\  
JKB~ 89.0 &   29.46 $\pm$   3.04 &    2.33 $\pm$   0.24 &   29.46 $\pm$   3.04 &    2.33 $\pm$   0.24    &  9.45 $\pm$  0.83 \\  
JKB~ 91.0 &    6.58 $\pm$   0.66 &    1.30 $\pm$   0.24 &    6.58 $\pm$   0.66 &    1.30 $\pm$   0.24    &  5.72 $\pm$  1.36 \\  
JKB~ 97.0 &    5.97 $\pm$   0.61 &    1.22 $\pm$   0.23 &    7.46 $\pm$   0.65 &    1.74 $\pm$   0.25    &  7.70 $\pm$  1.37 \\  
JKB~ 97.1 &    1.49 $\pm$   0.21 &    0.52 $\pm$   0.11 &                      &                         & $<$ 9.69 \\  
JKB~119.0$^\star$&    0.60 $\pm$   0.14 &    0.29 $\pm$   0.07 &    0.60 $\pm$   0.14 &    0.29 $\pm$   0.07    & $<$ 9.95 \\  
JKB~122.0$^\star$&    0.10 $\pm$   0.01 &    0.10 $\pm$   0.04 &    0.10 $\pm$   0.01 &    0.10 $\pm$   0.04    &  6.41 $\pm$  1.01 \\  
JKB~131.0 &    8.29 $\pm$   1.01 &    1.50 $\pm$   0.27 &    8.29 $\pm$   1.01 &    1.50 $\pm$   0.27    & $<$ 9.35 \\  
JKB~133.0 &    8.52 $\pm$   0.89 &    1.53 $\pm$   0.27 &   33.71 $\pm$   2.79 &    3.52 $\pm$   0.34    & $<$ 8.57 \\  
JKB~133.1 &   25.19 $\pm$   2.64 &    1.99 $\pm$   0.21 &                      &                         &  7.18 $\pm$  1.17 \\  
JKB~136.0 &   18.19 $\pm$   1.84 &    2.44 $\pm$   0.45 &   18.19 $\pm$   1.84 &    2.44 $\pm$   0.45    & $<$ 9.15 \\  
JKB~137.0 &  100.14 $\pm$  40.23 &    7.91 $\pm$   3.18 &  100.14 $\pm$  40.23 &    7.91 $\pm$   3.18    & $<$ 9.24 \\  
JKB~138.2 &   10.75 $\pm$   4.62 &    1.76 $\pm$   0.30 &   10.75 $\pm$   4.62 &    1.76 $\pm$   0.30    & $<$11.17 \\  
JKB~141.0 &  190.59 $\pm$  21.26 &   15.06 $\pm$   1.68 &  190.59 $\pm$  21.26 &   15.06 $\pm$   1.68    &  8.02 $\pm$  0.92 \\  
JKB~142.0$^\star$&    2.09 $\pm$   0.23 &    0.64 $\pm$   0.14 &    2.09 $\pm$   0.23 &    0.64 $\pm$   0.14    & $<$ 7.33 \\  
JKB~144.0 &  392.82 $\pm$  41.34 &   31.03 $\pm$   3.27 &  410.35 $\pm$  41.51 &   33.42 $\pm$   3.30    &  4.18 $\pm$  0.45 \\  
JKB~144.1 &   17.53 $\pm$   3.72 &    2.39 $\pm$   0.44 &                      &                         & $<$10.60 \\  
\hline
\end{tabular}
\label{tab:SFR}
\end{footnotesize}
\end{center}

\end{table*}

\subsubsection{Age of the current ionising population}
Following the methodology given in J15, we investigate the current age of the star-forming population via the luminosities of hydrogen recombination lines, \ha\ and \hb, which provide an estimate of the ionising flux present, assuming a radiation bounded nebula \citep{Schaerer:1998}. Using the Balmer line EW and its metallicity, we estimate the age of the latest star-forming episode in each object by comparing the EWs with those predicted by the spectral synthesis code, \textsc{starburst99} \citep{Leitherer:2010}. Details of the models can be found in J15. We interpolate between the models at the metallicity of each galaxy and assess the age of the current stellar population according ot its EW(\hb) and EW(\ha). The mean and half-difference between these two ages are given in Table~\ref{tab:SFR}.

The ages of the current ionising stellar populations within the BDD galaxies were found to range between 4.2 and 11.5~Myr, with a mean age of 7.4$\pm$1.9~Myr (this excludes galaxies for which lower limit oxygen abundances are available, as only upper limits can be calculated in these cases). These ages suggest that BDD galaxies have \hii\ regions either experiencing their first cycle of star-formation or undergoing a recent burst of star-formation.  With regards to Wolf-Rayet (WR) stars, which tend to exist between 3 and 6~Myr \citep{Leitherer:1999}, the WR feature at $\sim$4690~\AA\ is identified in only three objects; KJB 3, 7, and 119. Although this is somewhat surprising given the relatively young ages of the current ionising stellar population within BDDs, WR features are not typically seen in young XMP galaxies \citep{Shirazi:2012}. We do, however, detect the high-ionisation line \heii~$\lambda$4686 in five BDDs, a nebular line often seen in low metallicity galaxies. The mechanism(s) behind the existence of this line in nearby systems is still under debate and include WR stars, shocks, or X-ray binaries. In XMP galaxies such as IZw~18 and those observed here, \citet{Kehrig:2015} suggest that either low-metallicity super-massive O stars or rotating metal-free stars are more likely culprits for such hard ionising radiation. Since only one of the five BDD galaxies in which  \heii~$\lambda$4686 is detected (JKB~1, 7, 16, 22, and 29) is not an XMP galaxy, we cannot rule out that any of the proposed mechanisms for this line are at play within these objects.

\subsection{Photometric properties}\label{sec:phot}
Table~\ref{tab:photvals} lists the photometric SDSS $g$-band magnitudes for the BDD sample in Table~\ref{tab:gals}, along with the $u-g$,$g-r$,$r-i$, and $i-z$ colours. Total magnitudes were measured within the $R_{25}$ (the radius where the azimuthally averaged $r-$band surface brightness reaches 25~mag/arcsec$^2$), while the surface brightnesses were measured within the effective radii (radius with 50\%\ of the total flux). Also listed are the absolute $B$-band magnitudes computed using  distances listed from Table~\ref{tab:gals} \footnote{In order to convert the magnitude from the Sloan photometric system to Johnsons we used the prescription of Lupton 2005 \url{http://classic.sdss.org/dr7/algorithms/sdssUBVRITransform.html#Lupton2005}} and the $V$-band surface brightness, $\mu_v$. Uncertainties on all distance-dependent photometric properties resulting from distance uncertainties are expected to be $\lesssim0.5$~mag. For objects with $D>20$~Mpc, total $g$-band apparent magnitudes lie between 21.0 and 16.3~mag, absolute magnitudes are between $\sim-10 $ and $-18$~mag and surface brightnesses between 23 and 25 mag/arcsec$^{-2}$. Each of these are somewhat typical values for dwarf  galaxies although not as bright as many of the starbursting BCD galaxies. 
\begin{table}
\caption{Photometric properties of the BDD sample calculated from SDSS images. Objects without $M_b$ values are those for which spectroscopic redshifts could not be measured. Objects noted with `$^\star$' are those with uncertain distances, i.e. $D<20$~Mpc, and consequential photometric uncertainties of $>1$~mag.}
\begin{footnotesize}
\begin{center}
\begin{tabular}{|l|ccccccc|}
\hline
Obj ID & $g$ & $u-g$ & $g-r$ & $r-i$ & $i-z$ & $M_b$ & $\mu_v$\\
\hline
JKB~1& 17.32 &  0.92 &  0.08 & -0.13 & -0.42 &   -17.85&23.73\\
JKB~2& 18.11 &  0.87 &  0.19 &  0.14 &  0.15 &   -15.49&24.23\\
JKB~3& 17.05 &  0.80 &  0.23 &  0.07 &  0.25 &   -16.86&23.86\\
JKB~4$^\star$& 18.00 &  0.64 &  0.18 & -0.02 &  0.30 &   -11.39&23.89\\
JKB~5$^\star$& 17.95 &  0.87 &  0.24 &  0.10 & -0.76 &   -12.77&23.93\\
JKB~7$^\star$& 16.93 &  0.76 &  0.27 &  0.11 &  0.28 &   -12.89&23.68\\
JKB~8& 17.37 &  0.98 &  0.50 &  0.36 &  0.16 &   -17.67&23.92\\
JKB~10& 17.75 &  0.67 &  0.23 &  0.07 &  0.18 &  - &23.76\\
JKB~15& 18.14 &  1.19 & -0.59 &  0.15 & -0.71 &   -15.88&24.14\\
JKB~16& 17.72 &  0.78 &  0.17 &  0.10 & -0.01 &   -14.32&23.84\\
JKB~18$^\star$& 15.57 &  0.99 &  0.35 &  0.13 &  0.02 &   -15.33&22.79\\
JKB~19& 16.92 &  0.98 &  0.39 &  0.30 &  0.10 &  - &24.04\\
JKB~21& 18.59 &  0.85 &  0.34 &  0.30 & -0.02 &   -16.91&24.07\\
JKB~22& 17.45 &  0.67 &  0.13 &  0.01 &  0.11 &   -15.74&23.05\\
JKB~24& 16.51 &  0.88 &  0.36 &  0.17 &  0.07 &   -17.57&23.33\\
JKB~25& 17.66 &  0.81 &  0.27 &  0.43 &  0.44 &   -16.66&23.59\\
JKB~26& 18.78 &  1.22 &  0.26 &  0.15 &  0.10 &   -14.60&23.49\\
JKB~28& 18.31 &  0.61 &  0.11 & -0.26 &  0.31 &   -15.44&23.55\\
JKB~29& 17.59 &  0.67 &  0.12 & -0.06 &  0.47 &   -17.10&24.24\\
JKB~32& 17.03 &  0.90 &  0.49 &  0.25 &  0.03 &   -16.55&24.07\\
JKB~33& 18.87 &  0.54 &  0.29 &  0.08 &  0.23 &  - &24.09\\
JKB~37& 17.72 &  0.99 &  0.34 &  0.14 &  0.15 &   -15.82&23.79\\
JKB~40& 20.55 &  0.82 &  0.23 &  0.36 &  0.02 &   -12.84&24.71\\
JKB~42& 19.24 &  0.86 &  0.26 &  0.20 & -0.14 &   -16.09&24.50\\
JKB~53$^\star$& 16.95 &  0.82 &  0.18 &  0.06 & -0.06 &   -11.06&22.13\\
JKB~62& 18.03 &  0.65 &  0.11 &  0.01 &  0.13 &   -13.96&23.84\\
JKB~64$^\star$& 18.04 &  0.41 &  0.10 & -0.24 &  0.02 &   -11.41&23.33\\
JKB~69& 18.79 &  0.50 &  0.06 &  0.04 &  0.34 &   -14.92&23.37\\
JKB~74& 19.12 &  0.96 &  0.39 &  0.24 &  0.13 &  - &24.28\\
JKB~78$^\star$& 18.64 &  0.62 &  0.13 &  0.01 &  0.15 &   -10.25&24.25\\
JKB~80$^\star$& 17.41 &  0.77 &  0.22 &  0.20 &  0.35 &   -12.44&24.05\\
JKB~83$^{\star\dagger}$& 20.00 & -0.22 &  0.50 & -0.88 &  0.16 &    -4.10&23.26\\
JKB~85$^\star$& 18.76 &  0.64 &  0.11 & -0.11 & -0.06 &   -15.96&24.11\\
JKB~89& 18.02 &  0.79 &  0.14 &  0.08 &  0.19 &   -15.98&23.10\\
JKB~91& 18.58 &  0.85 &  0.26 &  0.09 &  0.14 &   -12.81&22.75\\
JKB~97& 18.34 &  0.73 &  0.10 & -0.02 & -0.00 &   -14.13&24.00\\
JKB~102& 19.70 &  1.34 &  0.66 &  0.21 & -0.39 &  - &24.52\\
JKB~107& 20.99 & - &  0.69 &  0.45 &  0.61 &  - &24.75\\
JKB~114& 17.79 &  0.74 &  0.14 &  0.16 & -0.02 &  - &23.68\\
JKB~119$^\star$& 18.34 &  0.96 &  0.44 &  0.24 &  0.15 &   -11.95&23.63\\
JKB~122$^\star$& 16.32 &  0.81 &  0.27 &  0.33 & -0.75 &    -9.37&23.07\\
JKB~124& 17.78 &  0.87 &  0.51 &  0.31 &  0.22 &  - &24.04\\
JKB~129& 17.03 &  0.85 &  0.33 &  0.18 &  0.01 &  - &23.32\\
JKB~131& 18.06 &  0.79 &  0.32 &  0.10 &  0.24 &   -14.79&23.42\\
JKB~133& 17.58 &  1.20 &  0.44 &  0.25 &  0.26 &   -15.85&23.27\\
JKB~136& 17.86 &  0.90 &  0.31 &  0.09 &  0.18 &   -16.20&23.45\\
JKB~137& 16.31 &  1.11 &  0.52 &  0.22 &  0.14 &   -17.58&22.90\\
JKB~138& 17.28 &  1.09 &  0.53 &  0.30 &  0.04 &   -15.93&23.63\\
JKB~141& 17.23 &  1.04 &  0.18 &  0.14 &  0.45 &   -17.01&24.09\\
JKB~142$^\star$& 18.08 &  0.81 &  0.20 &  0.08 &  0.02 &   -11.60&23.41\\
JKB~144& 17.41 &  1.17 &  0.21 &  0.08 &  0.36 &   -17.28&23.64\\
\hline
\end{tabular}
\end{center}
\label{tab:photvals}
$^\dagger$ Known to be a single \hii\ region within M~81.
\end{footnotesize}

\end{table}

\begin{table}
\caption{Stellar masses derived from stellar population synthesis and SDSS colours listed in Table~\ref{tab:photvals}. Also listed are SSFRs derived from $M_\star$ and total SFRs listed in Table~\ref{tab:SFR}.  Details can be found in Section~\ref{sec:mass}. Objects noted with `$^\star$' are those with uncertain distances, i.e. $D<20$~Mpc.}
\begin{footnotesize}
\begin{center}
\begin{tabular}{|l|cc|}
\hline
Obj ID & log($M_\star$)/$M_\odot$ & log(SSFR/yr$^{-1}$) \\
\hline
JKB~1 &  8.10$\pm$ 0.42 &  -9.10$\pm$  0.42 \\
JKB~2 &  7.80$\pm$ 0.32 &  -9.15$\pm$  0.33 \\
JKB~3 &  8.34$\pm$ 0.29 &  -9.52$\pm$  0.29 \\
JKB~4$^\star$ &  5.91$\pm$ 0.32 &  -8.39$\pm$  0.32 \\
JKB~5$^\star$ &  6.30$\pm$ 0.48 &  -8.67$\pm$  0.48 \\
JKB~7$^\star$ &  6.82$\pm$ 0.27 &  -9.22$\pm$  0.27 \\
JKB~8 &  8.95$\pm$ 0.25 &  -9.60$\pm$  0.25 \\
JKB~15 &  6.65$\pm$ 0.05 &  -7.58$\pm$  0.05 \\
JKB~16 &  7.21$\pm$ 0.32 &  -9.02$\pm$  0.32 \\
JKB~18$^\star$ &  7.95$\pm$ 0.19 & -10.08$\pm$  0.19 \\
JKB~22 &  7.60$\pm$ 0.28 &  -8.89$\pm$  0.28 \\
JKB~24 &  8.80$\pm$ 0.24 &  -9.50$\pm$  0.24 \\
JKB~25 &  8.29$\pm$ 0.27 &  -9.57$\pm$  0.27 \\
JKB~26 &  7.66$\pm$ 0.18 &  -9.18$\pm$  0.18 \\
JKB~28 &  7.24$\pm$ 0.34 &  -8.91$\pm$  0.34 \\
JKB~29 &  8.12$\pm$ 0.29 &  -9.41$\pm$  0.29 \\
JKB~37 &  8.15$\pm$ 0.20 &  -9.79$\pm$  0.20 \\
JKB~40 &  6.82$\pm$ 0.27 &  -8.76$\pm$  0.27 \\
JKB~42 &  8.10$\pm$ 0.23 &  -8.84$\pm$  0.24 \\
JKB~53$^\star$ &  6.03$\pm$ 0.23 &  -8.63$\pm$  0.24 \\
JKB~62 &  6.89$\pm$ 0.28 &  -8.85$\pm$  0.28 \\
JKB~64$^\star$ &  5.47$\pm$ 0.27 &  -8.02$\pm$  0.27 \\
JKB~69 &  7.04$\pm$ 0.28 &  -8.53$\pm$  0.29 \\
JKB~78$^\star$ &  5.34$\pm$ 0.36 &  -7.97$\pm$  0.36 \\
JKB~83$^{\star\dagger}$ &  2.78$\pm$ 0.13 &  -6.48$\pm$  0.13 \\
JKB~85$^\star$ &  7.51$\pm$ 0.27 &  -9.05$\pm$  0.27 \\
JKB~89 &  7.90$\pm$ 0.26 &  -9.54$\pm$  0.26 \\
JKB~91 &  6.84$\pm$ 0.23 &  -8.73$\pm$  0.23 \\
JKB~97 &  7.03$\pm$ 0.25 &  -8.79$\pm$  0.25 \\
JKB~119$^\star$ &  6.69$\pm$ 0.25 &  -9.23$\pm$  0.25 \\
JKB~122$^\star$ &  5.38$\pm$ 0.23 &  -8.38$\pm$  0.23 \\
JKB~131 &  7.59$\pm$ 0.28 &  -9.41$\pm$  0.28 \\
JKB~133 &  8.33$\pm$ 0.21 &  -9.79$\pm$  0.21 \\
JKB~136 &  8.19$\pm$ 0.25 &  -9.81$\pm$  0.25 \\
JKB~137 &  9.02$\pm$ 0.21 & -10.12$\pm$  0.21 \\
JKB~138 &  8.31$\pm$ 0.25 & -10.06$\pm$  0.25 \\
JKB~141 &  8.52$\pm$ 0.23 &  -9.35$\pm$  0.23 \\
JKB~142$^\star$ &  6.23$\pm$ 0.25 &  -8.42$\pm$  0.25 \\
JKB~144 &  8.69$\pm$ 0.20 &  -9.17$\pm$  0.20 \\
\hline
\end{tabular}
\label{tab:mass}
\end{center}
$^\dagger$ Known to be a single \hii\ region within M~81.
\end{footnotesize}

\end{table}

\subsubsection{Stellar masses}\label{sec:mass}
Stellar masses ($M_\star$) were estimated by modelling the BDDs stellar content via a flexible stellar population synthesis code \citep[FSPS,][]{Conroy:2009} using the broad-band colours listed in Table~\ref{tab:photvals}.  A grid of synthetic spectra was created assuming a constant star-formation history, a Chabrier IMF and the Milky Way extinction law from \citet{Cardelli:1989}.  Each spectrum was synthesised for a range of stellar metallicities ($0.01 <Z_\ast/\Zsol<1.5$) and $V$-band optical depths of $\tau_v=$0--2 for a dust screen.  Synthetic SDSS colours were computed from each spectrum at the correct redshift of each object as a function of $Z_\ast,\tau_v,$ and age.  Because the FSPS models can be evaluated at arbitrary points within the bounds of the cube, we adopt the following priors on  
$Z_\ast$, $\tau_v,$ and age: 
$Z_\ast = Z_{\rm H\,\textsc{ii}} \pm 0.3$\,dex, 
$-4<\log(\tau_v)<0.3$, and age$> 1$\,Myr, respectively.

For each object the likelihood of the colours given in Table~\ref{tab:photvals} and metallicities given in Table~\ref{tab:abunds} is sampled using a Markov chain Monte Carlo methodology, producing a posterior probability distribution for the parameters of the model. We do not correct the broad-band magnitudes for the contribution of emission lines because the effect is very minor (typically $ \lesssim0.1$~mag) so can usually be ignored. Whilst we are aware that a constant star-formation history may offer an over-simplified description of our systems, without any additional information from their resolved stellar populations, we are not able to adopt a more physical model. 

The resultant stellar masses for our sample of BDDs are listed in Table~\ref{tab:mass}. Uncertainties on each mass are dominated by the uncertainties on the photometric measurements and possible offsets between $Z_\ast$ and $Z_{\hii}$. We expect uncertainties on the stellar mass due to distance uncertainties to be $\lesssim0.3$~dex. Stellar masses are in the range  $\log(M_\star/\Msol) \sim $5-9, with an average of $\log(M_\star/\Msol) =7.3\pm1.0$ (these numbers exclude JKB~83 which is known to be a single \hii\ region in M81).  Such masses are typical of BCD and irregular dwarf galaxies, albeit towards the low-mass end.  It should be noted here that there are several caveats hindering our stellar mass calculations. Firstly, we are relying on SDSS photometry and only computing total fluxes where the surface brightness of galaxies is above ($<\mu_v>\sim25$\,mag/arcsec$^2$), which may lead us to lose some stellar mass in the very diffuse stellar component. Secondly, significant uncertainties in stellar mass estimates can exist due the assumed constant star-formation history. Finally, we lack infrared imaging \citep[e.g.,][]{Zhao:2013}, which would make our stellar mast estimates more robust. However, in the absence of deeper optical or infrared imaging, we adopt the values calculated here and restrict our mass-related analysis to the mass-metallicity relation described in the subsequent section, for the sake of comparison with other samples. In the future, we hope to obtain deep broad-band optical and infrared imaging of the entire sample and undertake a detailed and accurate photometric analysis of these low-luminosity objects.
\begin{figure*}
\includegraphics[scale=0.5,angle=90]{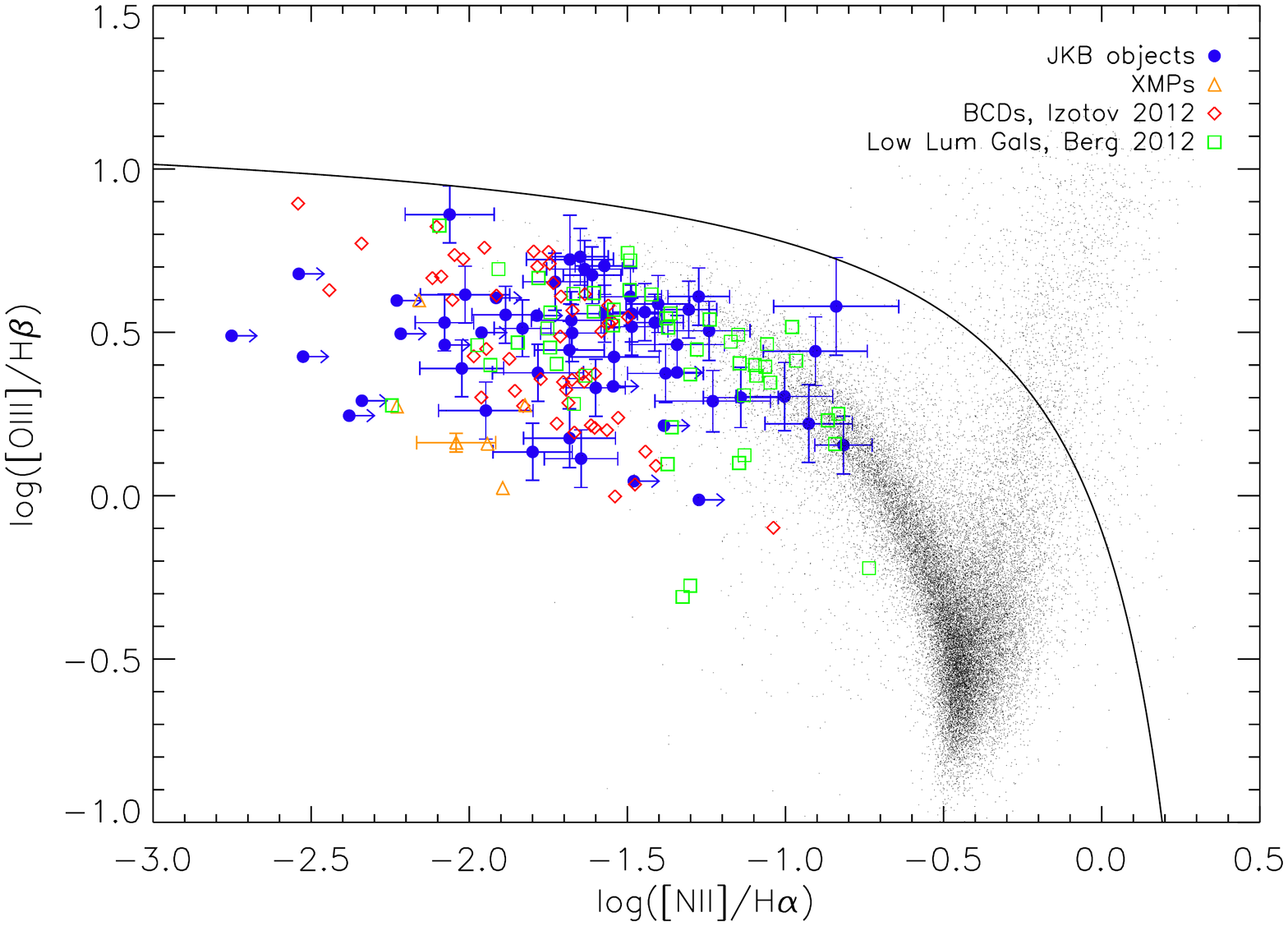}
\caption{The `BPT' emission line diagnostic diagram, showing \foiii ($\lambda5007$)/\hb\, vs. \fnii($\lambda6584$)/\ha, traditionally used to separate galaxies according to their metallicity, and strength and hardness of their ionising radiation.  In black we show the SDSS emission line galaxies (selected according to the criteria of \citet{Brinchmann:2008}), which form a branch of star-forming galaxies on the left, and Seyferts/LINERs on the right.  The BDD galaxies (i.e., JKB~objects listed in Table~\ref{tab:gals}) are distributed throughout the top-left region of the star-forming branch where galaxies of low metallicity and high ionisation fall.   For comparison, we also plot a selection of XMPs, BCDs and low-luminosity dwarf galaxies (see text for details).  The black solid line represents the `maximum starburst line' of \citep{Kewley:2001}. }
\label{fig:BPT}
\end{figure*}

Table~\ref{tab:mass}  also lists the specific star-formation rates (SSFR=SFR/$M_\star$) for each galaxy, calculated from the total SFR values (Table~\ref{tab:SFR}) and stellar masses.  SSFR are in the range $-10.0<\log($SSFR/yr$^{-1}$)$<-6.5$, with a mean $\log($SSFR/yr$^{-1})=-9.0\pm0.7$.  Such SSFRs are typical of low mass dwarf galaxies \citep[e.g.,][]{Lee:2009,Huang:2012} and, with respect to their stellar masses, fit nicely within the narrow SSFR range seen within the low-mass range of the `star-forming sequence' \citep{Schiminovich:2007}.


\section{Discussion}\label{sec:disc}

The conclusion of J15, which considered a sample of only 12 JKB objects, was that blue diffuse dwarfs have properties reminiscent of both dwarf irregular galaxies and blue compact dwarf galaxies. In fact, in consideration of their star-formation rates and surface brightness fluxes, they appear to lie directly in between these two populations. In this section, we assess whether these conclusions still hold now that we have increased our spectroscopic sample of JKB objects by a factor of four. 

In each of the following sections, we compare the JKB objects against a variety of dwarf galaxy samples. We use the comparison sample adopted in J15 along with some additional dIrr and blue compact dwarf galaxy samples. Overall this includes: XMP galaxies [Leo~P \citep{Skillman:2013}, LeoA \citep{VanZee:2006}, UGCA292 \citep{VanZee:2000}, DDO68 \citep{Pustilnik:2005}, SBS1129+576 \citep{Ekta:2006,Guseva:2003}, SBS1129+577 \citep{Ekta:2006}, J2104-0035 and UGC772 \citep{Ekta:2008}, UM133 and SDSSJ011914 \citep{Ekta:2010}, HS2134 \citep{Pustilnik:2006}] and the sample of \citet{Morales-Luis:2011};  low-luminosity and dwarf irregular galaxies of \citet{Berg:2012}, \citet{Haurberg:2013} and the SHIELD galaxies from \citet{Haurberg:2013}; and blue compact dwarfs (BCDs) from \citet{Izotov:2007}, \citet{Izotov:2012}, and \citet{Berg:2016}.  Within each of the figures, the three groups of dwarf galaxies (XMP galaxies, low luminosity dwarf irregulars, and BCDs) are colour-coded orange, green, and red, respectively.  In order to avoid discrepancies between metallicity measurements, we only include galaxies for which a direct measurement of the metallicity has been reported.  To ensure a fair comparison with our JKB objects, all SFRs for the comparison samples have been calculated according to the method described above, i.e., utilising the \citet{Kennicutt:1998} relation for all galaxies with $L$(\ha)$>3.5\times10^{39}$~erg\,s$^{-1}$ and the \citet{Lee:2009} recalibration for those with luminosities below this cut-off. 
\begin{figure*}
\includegraphics[scale=0.5,angle=90]{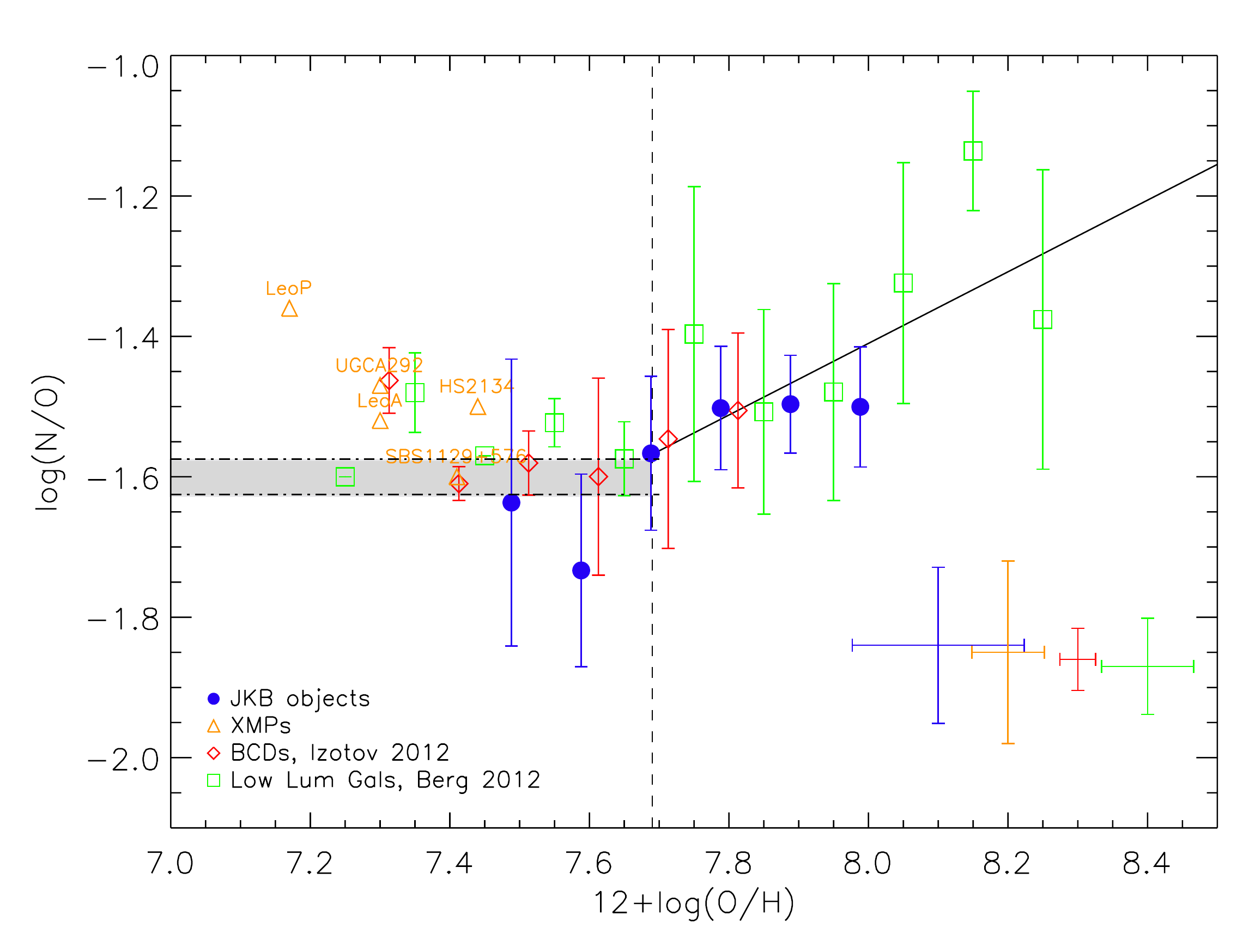}
\caption{The relationship between oxygen abundance and nitrogen-to-oxygen ratio in the low-metallicity regime. We show the results from our sample of BDD galaxies (i.e., JKB~objects listed in Table~\ref{tab:gals}, blue circles), where metallicity and N/O values correspond to the average measured for that system. For comparison we plot the recently compiled sample of BCDs by \citet{Izotov:2012} and the low-luminosity dwarf galaxies of \citet{Berg:2012}. Each sample has been binned by $\Delta\log$(O/H)=0.1 and error bars on each point represent the standard deviation within that bin. We additionally show the individual values from a selection of XMP galaxies (see text for details). Error bars at the bottom right of the plot represent the typical uncertainty on a single measurement within that sample.
The black solid line shows the relationship between log(N/O) and log(O/H) in the metallicity range $12+\log{\rm (O/H)} > 7.7$ derived by \citet{Berg:2012} for low-luminosity galaxies, while the grey shaded region is the narrow N/O plateau proposed for XMPs by \citet{Izotov:1999}. The vertical dashed line represents the XMP cut-off (12+log(O/H)$<$7.69).} 
\label{fig:met_no}
\end{figure*}

\subsection{Emission line diagnostics}
If we first look to Figure~\ref{fig:BPT}, we place our new, larger sample of BDDs onto the traditional emission-line diagnostic plot, log{\foiii/\hb} vs. log(\fnii/\ha) \citep[i.e., the traditional `BPT' diagram introduced by][]{Baldwin:1981}.  The BDDs are distributed within the low-metallicity, high-ionisation parameter (i.e., $U$-parameter, defined as the number density ratio of ionising photons to particles) region of the diagram, alongside the comparison sample of BCDs, XMPs and low luminosity galaxies. Whilst we do not list the $U$-parameter for individual galaxies, they were calculated following the method outlined in J15 and found to have the range $-3.59<\log(U)<-2.19$. Such properties are to be expected, given the low-metallicity, highly ionised gas surrounding their young photoionising stellar populations. 

A distinct offset can be seen in Fig.~\ref{fig:BPT} between SDSS galaxies and low-metallicity dwarf galaxies.  This is due to low-metallicity systems having smaller \foiii/\hb\ and \fnii/\ha\ ratios than high-metallicity systems, as noted in \citet{Izotov:2012}. Also, SDSS galaxies do not typically occupy this region due to the rarity of low-metallicity galaxies. 
\begin{figure*}
\includegraphics[scale=0.5,angle=90]{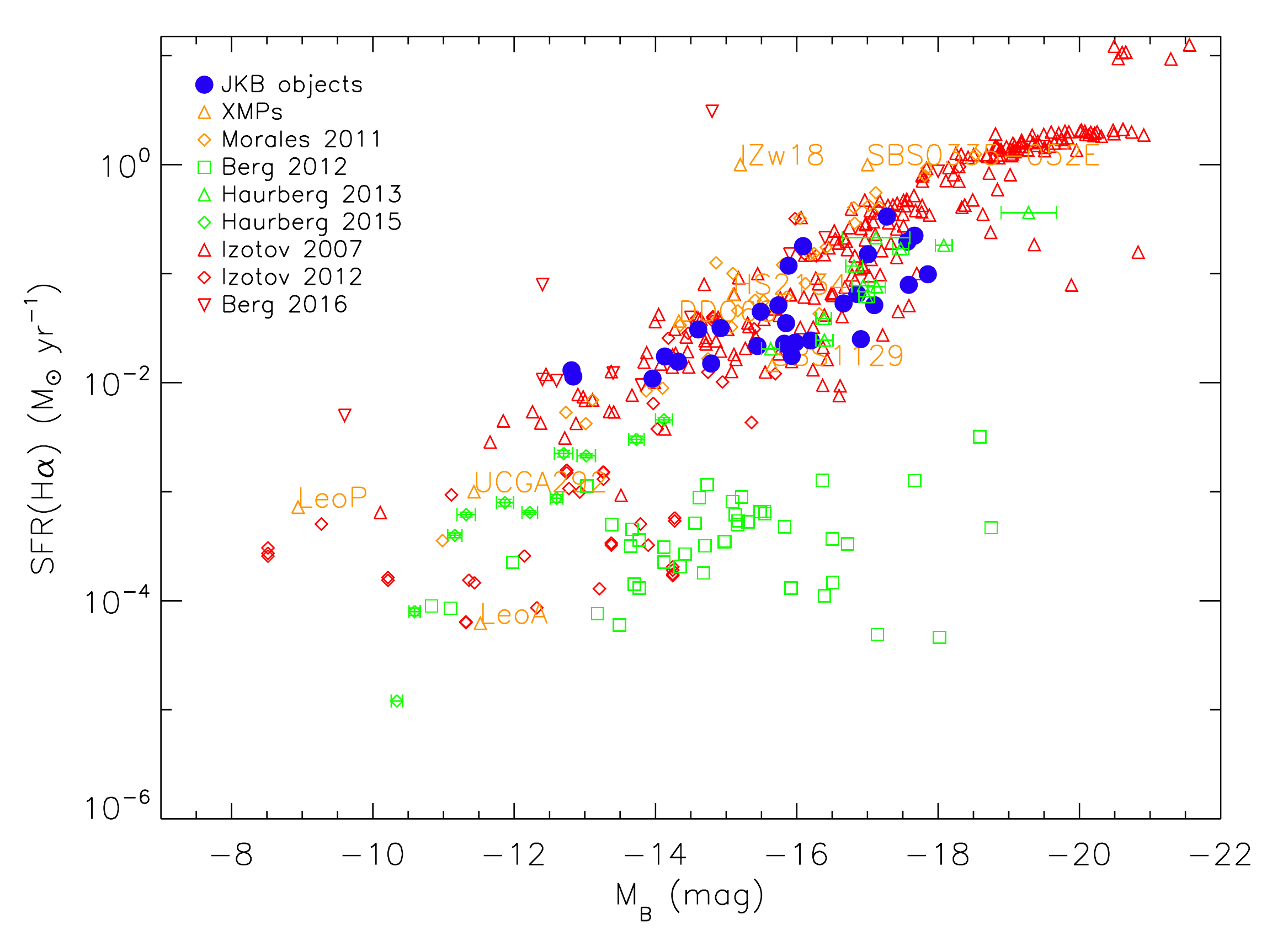}
\caption{The relationship between absolute $B$-band magnitude ($M_B$) and total star-formation rates for our sample of blue diffuse dwarf galaxies (i.e., JKB~objects listed in Table~\ref{tab:gals}). We also overplot the comparison sample described in Section~\ref{sec:disc}.}
\label{fig:Mb_SFR}
\end{figure*}

\subsection{Nitrogen-to-oxygen ratio}
With regards to their chemical properties, JKB objects have metallicities in the range 7.43$<$12+log(O/H)$<$8.01, i.e., 0.05--0.21~(O/H)$_\odot$. Nitrogen-to-oxygen ratios were found to lie in the range $-1.88<$log(N/O)$<-1.39$ and are typical values for the metallicity range of the sample.  This is evident in Figure~\ref{fig:met_no}, where we show the distribution of $\log$(N/O) as a function of metallicity for our sample (excluding those with N/O limits) alongside other low-metallicity galaxies such as BCDs \citep{Izotov:2012}, low luminosity galaxies from \citet{Berg:2012}, and other known XMP galaxies.  
In J15 we were unable to make any firm conclusions on the reality of the scatter in N/O values at low metallicity due to our limited sample size and limited accuracy of abundance determinations.  Here, with the N/O values from 26 spectra, we can see that while a plateau may exist in N/O for XMP objects, the relatively large scatter in N/O for galaxies within this metallicity regime suggests that the plateau may not be as narrow as proposed by \citet{Izotov:1999}. For example, the galaxies in Fig.~\ref{fig:met_no} with 12+log(O/H)$<$7.69 have a mean N/O value of 
$\langle\log\textrm{(N/O)}\rangle=-1.55\pm0.02$ with an intrinsic 1~dex scatter of $0.10$.  

In contrast to the BDD galaxies and other typical low-mass dwarf galaxies, some XMPs and BCDs have been found to have high N/O values for their metallicity. For example, \citet{Sanchez-Almeida:2016} find that 10\%\ of the 332 XMPs within their sample have log(N/O)$\gtrsim-1.2$ \footnote{We do not include this sample in Fig.~\ref{fig:met_no} because chemical abundances were not calculated using the direct-method}. Such large N/O values at low O/H can be attributed to the accretion of metal-poor gas - i.e., mixing which lowers the abundance of O whilst maintaining the original N/O ratio \citep[see e.g.,][ and references therein]{Amorin:2010}. If this is indeed the case, the relatively normal N/O ratio in our BDD sample suggests that the accretion of significant amounts of metal-poor gas does not appear to be at play within BDDs, as we will see in subsequent sections. However, given our small sample size (i.e., chemical abundance measurements for only 26 JKB objects) and a probable uncertainty on the expected 10\%\ fraction showing high N/O values from \citet{Sanchez-Almeida:2016}, we cannot rule out that high N/O values would indeed be seen in larger sample.  In support of this conclusion, it should also be noted that the large scatter in N/O at intermediate metallicities is not solely restricted to XMP and BCD galaxies. \mbox{\citet{VanZee:2006}} found that a large scatter in N/O also exists for isolated dwarf irregular galaxies with 12+log(O/H)$<$8.2, which they attribute to either varied star formation histories or an increased production of secondary nitrogen at intermediate metallicities.


\subsection{Combined Properties}\label{sec:combi}

The most efficient way of assessing the placement of BDDs within the context of other dwarf galaxy populations is by combining their physical (e.g., SFR), chemical (e.g., metallicity), and photometric properties (e.g., magnitudes, masses). 

Firstly in Figure~\ref{fig:Mb_SFR} we plot star formation rate vs. B-band absolute magnitude . Despite the SFR recalibration towards higher SFRs for low luminosity galaxies (which was not implemented in J15), we can see that JKB objects still appear to lie mid-range in SFR between dwarf irregulars and BCDs, i.e., with the majority of JKB objects lying between $-12<M_B<-18$ and occupying the upper range of the dIrr SFR and the lower range of the BCD SFR distributions. Interestingly, the emission-line selected sample of XMP galaxies from \citet{Morales-Luis:2011} have SFRs that are similar to those of JKB objects, despite the fact that the JKB object's \hii\ regions were not bright enough to be selected for the SDSS spectroscopic sample. By design, \citet{Morales-Luis:2011} XMP sample are bursty or `active' XMPs, with mostly cometary morphologies, and therefore do not include the diffuse XMPs like Leo~P or those seen within our sample. 

\begin{figure*}
\includegraphics[scale=0.5,angle=90]{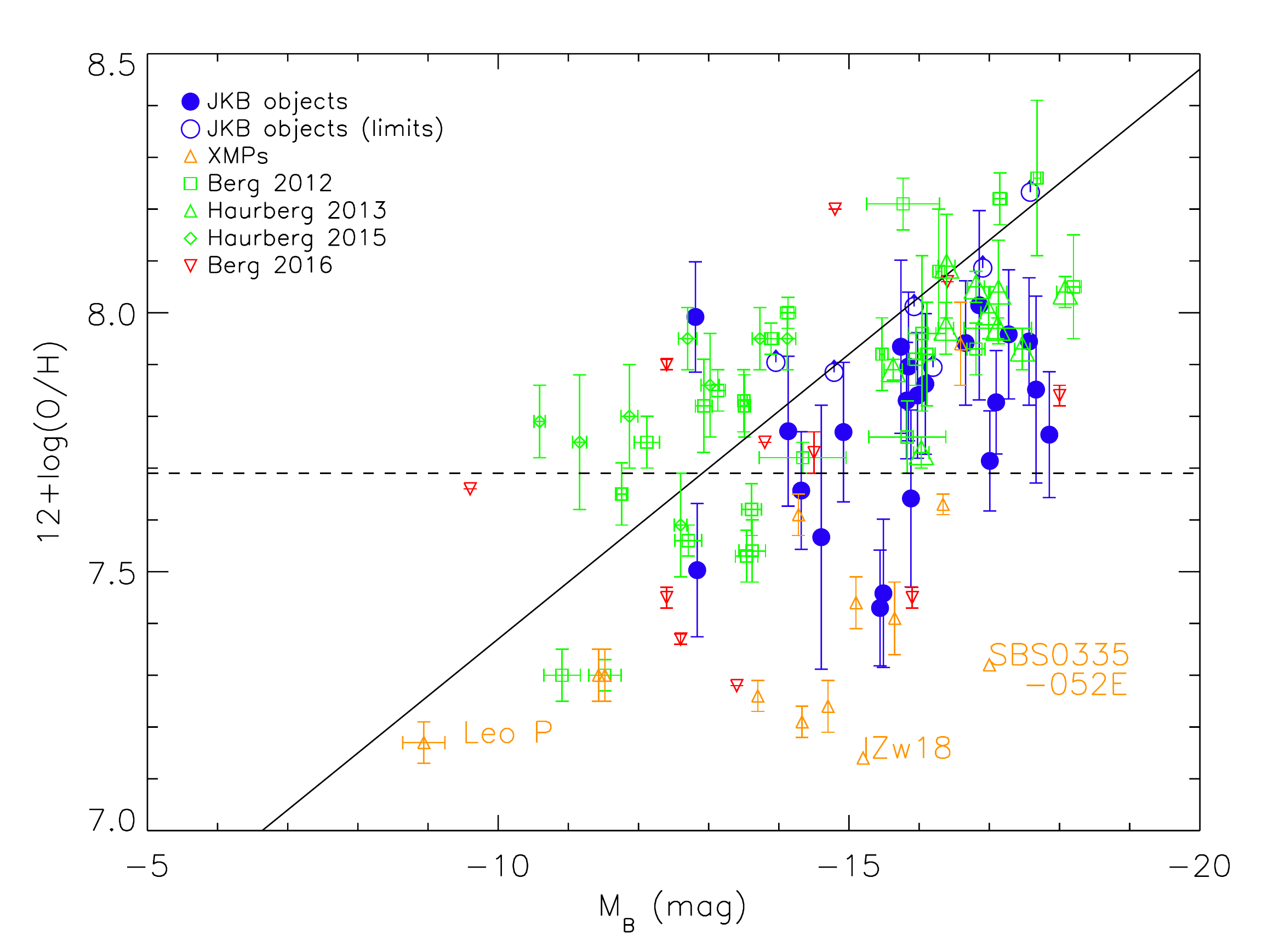}
\caption{The relationship between oxygen abundance and absolute $B$-band magnitude ($M_B$)  for our sample of blue diffuse dwarf galaxies (i.e., JKB~objects listed in Table~\ref{tab:gals}). In cases where multiple \hii\ regions exist, we plot the average O/H abundance measured within that system.  We also overplot the comparison sample described in Section~\ref{sec:disc} and the $L-Z$ relationship of \citet{Berg:2012}.}
\label{fig:mb_met}
\end{figure*}
Secondly, in Figure~\ref{fig:mb_met} we plot metallicity as a function of B-band magnitude, commonly known as the $L-Z$ relation. JKB~objects show a significant amount of scatter around and slightly below the $L-Z$ relation defined from the low-luminosity sample of \citet{Berg:2012} but generally are in agreement with low-luminosity dwarf irregular galaxies. They do not, however, show the same overall trend of XMP galaxies, who typically fall well below the relation. This is thought to be due to the inflow of pristine gas which both depletes the metals within the ISM and fuels star-formation within their \hii\ regions. 

\begin{figure*}
\includegraphics[scale=0.6,angle=90]{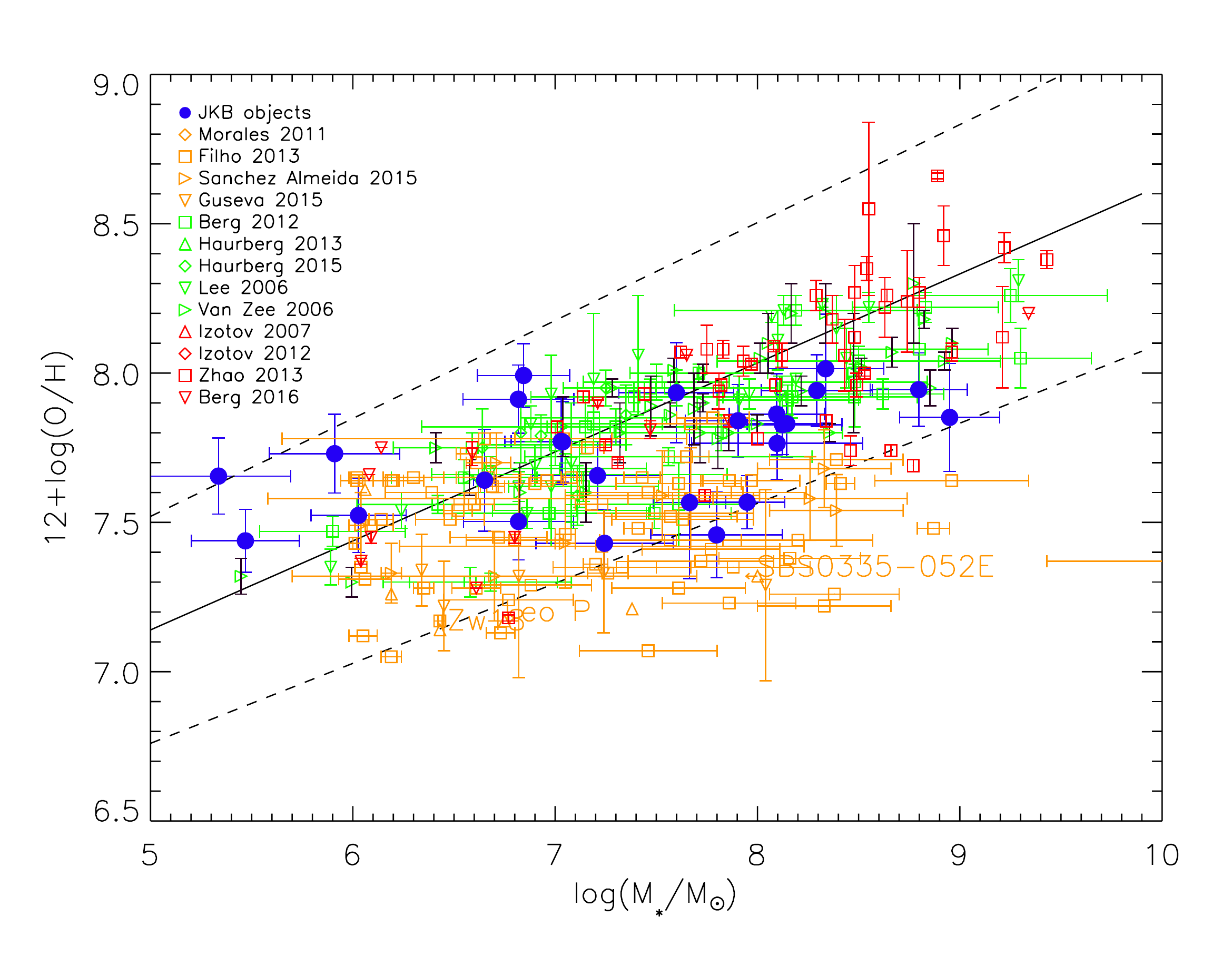}
\caption{The relationship between oxygen abundance and stellar mass for our sample of blue diffuse dwarf galaxies (i.e., JKB~objects listed in Table~\ref{tab:gals}). In cases where multiple \hii\ regions exist, we plot the average O/H abundance measured within that system. We overplot M-Z relation for low-mass galaxies derived by Lee et al. (2006) and the comparison sample described in Sections~\ref{sec:disc} and \ref{sec:combi}. It should be noted that stellar masses have additional, unquantified uncertainties owing to a presumed constant SFR model.}
\label{fig:mass_met}
\end{figure*}

A similar trend can be seen in Figure~\ref{fig:mass_met}, where we place the BDD galaxies into context with other dIrrs and BCDs within the mass-metallicity diagram.  Since values of mass and metallicity are more easily available in the literature than individual emission line fluxes,  here we can make several additions to our comparison sample: dIrrs from \citet{Lee:2006} and \citet{VanZee:2006}, BCD galaxies from \citet{Zhao:2013}, and a sample of XMP galaxies from \citet{Filho:2015}, \citet{Sanchez-Almeida:2015} and \citet{Guseva:2015}. Again, the XMPs tend to fall below the M-Z relation, which studies attribute to two possible scenarios: (i) the infall of metal-poor gas \citep[diluting the metal-content of the ISM and increasing the mass; e.g.,][]{Peeples:2009}; or (ii) they are genuinely young systems that are currently chemically unevolved \citep{Guseva:2015}. Since all XMP galaxies whose stellar populations are resolvable by the \textit{Hubble Space Telescope} have been found to have underlying populations $>1$~Gyr old \citep{Aloisi:1999,Aloisi:2007,Izotov:2002}, we consider the latter scenario unlikely. With regards to the former, given that the mass-metallicity relationship itself may result from gas accretion \citep[e.g.,][]{Mannucci:2010,Yates:2012} we cannot rule out the inflow of metal-poor gas completely. Instead, the fact that BDDs appear to lie within the low-mass M-Z relation perhaps suggests that they do not sustain excessive inflows of metal-poor gas for their stellar mass.



\subsection{Classifying blue diffuse dwarf galaxies}
Overall it can be seen that several properties of BDD galaxies overlap with those of other dwarf galaxy classes. With regards to their chemical properties, they are certainly low-metallicity objects. They have blue optical colours, owing to the fact that they host young photoionising stellar populations within their \hii\ regions. Whilst both these properties are typical of blue compact dwarf galaxies and other XMP galaxies, their morphologies are not.  The large majority of XMP galaxies are cometary in shape \citep{Papaderos:2008,Filho:2013,Sanchez-Almeida:2016}, with 
$\lesssim 10\%$ showing multiple knots of star formation. \citet{Skillman:2013} classify these cometary XMP-types as \textit{bursty}-XMPs, unlike Leo~P which is a \textit{quiescent}-XMP. Within this classification scheme, `quiescent' is associated with `diffuse' and `bursty' with `compact'.  Blue compact dwarf galaxies are very similar to \textit{bursty+compact} XMPs in this respect, with a large and luminous concentration of star-formation at their centre \citep[e.g.,][ and references therein]{GildePaz:2003}. While the `diffuse' aspect of BDDs aligns them closely with dIrrs in a morphological context, as we have seen from Fig~\ref{fig:Mb_SFR} their star-forming regions are more active than dIrrs. In this sense, the XMP galaxies within our sample of BDDs would be exceptions to the \citet{Skillman:2013} classification, as they are \textit{active+diffuse}-XMPs. 

In general, the larger population of BDDs appear to be a population of \textit{active} dwarf irregular galaxies that can extend down to extremely low metallicity, suggesting that the placement of BDDs within the grand scheme of dwarf galaxy evolution is not clear-cut.  In light of this `active dIrr' scenario, a question exists as to whether BDDs could offer an evolutionary link between dwarf irregular galaxies and blue compact dwarfs. The existence of such a link between these two subclasses of dwarf galaxies has long been a subject of debate. Originally it was thought that dIrrs were post-starburst BCDs \citep{Thuan:1985,Davies:1988}; however, in-depth studies of their surface brightness profiles suggest otherwise. For example, \citet{Janowiecki:2014} show that the underlying hosts of BCDs have significantly more concentrated light distributions, with properties that are so distinct from dIrrs that major structural changes would be required in order transform the two populations into one another. However, even if BDDs may not have been BCD galaxies in the past, they may represent a different phase of dIrr evolution when star-formation is more prevalent.  In order to determine the possible role of BDDs with respect to their neighbouring dwarf galaxy classifications (dIrrs and BCDs) we would require deep imaging of our sample - a definite future goal of this study.

However, what we \textit{can} say is that BDDs belong within the class of low-surface brightness (LSB) dwarf galaxies, a broad classification that encompasses both irregular and spheroidal dwarf galaxies with a central $B$-band surface-brightness $\mu_{B,0}\gtrsim$23~mag/arcsec$^{-1}$ \citep{Kunth:2000}. In particular, they would be classed as LSB dIrr galaxies. In this respect, we may have uncovered a sample of low-surface brightness dIrr galaxies that  bridge the BCD and dIrr populations with regards to physical and morphological properties - a population of actively star-forming galaxies where the \hii\ regions are not centrally clustered. 

The method by which star-formation occurs in BDDs could be the element that separates them from the accretion-induced star-formation witnessed in BCDs and XMPs (as demonstrated by the mass-metallicity relation) and the lack of star-formation seen in dIrrs. Despite the apparent absence of significant amounts of gas flowing into central star-forming complexes, they continue to host \hii\ regions embedded within low-luminosity, metal-poor gas. Whilst the star-formation could be linked to past/ongoing interactions, these galaxies were selected on the basis of being isolated, implying that only past interactions can be relevant. Alternatively, stars may be forming from previously ejected and reprocessed gas that has now cooled and fallen back onto the \hii\ regions.

The chemically-young nature of the gas within the \hii\ regions suggests very low-levels of star-formation in the past. This could be verified by resolving their stellar populations (for the closest BDDs) and mapping their star-formation histories (SFHs). We know that dwarf galaxies appear to have SFHs that suggest `gasping' and not bursting star-formation - i.e., long, moderate activity and short quiescent phases \citep{Annibali:2003} and dIrrs and BCDs have similar SFHs despite having very different current star-formation rates. As such, it would be interesting to determine whether BDDs have SFHs that are intermediate between these two classes, as their physical and chemical properties suggest. As such, the SFHs of BDDs would not only provide significant insight into their chemical evolutionary history but also their relation to other dwarf galaxy populations. On the other hand, integral-field-unit observations would enable us to assess the kinematic stability of the gas and assess the possible role of past interactions/mergers in their star-forming activity. Consolidating SFHs and kinematical maps will provide additional insight into whether the low metal-content of BDDs is environmentally driven or a natural consequence of evolution. For example, \citet{McQuinn:2015b} use SFHs and chemical evolution models to suggest that Leo~P's extremely low metal content is due to the combined effects of inefficient star-formation and it losing 95~\%\ of its oxygen via stellar feedback (e.g., outflows). It would be interesting to explore whether BDDs fall into the class of XMP galaxies that show signs of interaction \citep[e.g., the higher luminosity, low-mass XMPs of ][]{Izotov:2012} or the secularly evolving, low-mass, low-luminosity XMPs like Leo~P.  Such observations will form part of a future study using VLT-MUSE observations of a sub-sample of the BDD population (James et al., in-prep).

\section{Conclusions}\label{sec:conc}
In order to unveil more XMP objects \citet{James:2015} utilised the morphological properties of Leo~P to conduct a search within the SDSS imaging database. This resulted in $\sim$150 previously unstudied blue, diffuse galaxies with \hii\ regions scattered throughout low surface brightness gas, which we called  `blue diffuse dwarf' (BDD) galaxies. \citet{James:2015} concentrated on 12 BDD galaxies (or JKB objects) for which we had MMT optical spectroscopic data at the time.  Here we have extended this sample to 51 BDD galaxies in order to obtain a fuller picture of this population of dwarf galaxies.

The 51 galaxies in our sample are at distances of 1--150~Mpc (although we expect distance measurements below $20$~Mpc to be highly affected by peculiar motions) and are 20--50~arcsec across.  All of the spectra showed bright emission lines typical of star-forming galaxies, with the exception of 11 objects, which were not included in the spectroscopic analysis. These non-emission line objects may represent BDD galaxies observed during quiescent periods between star-formation episodes.  A full spectroscopic analysis was undertaken on 40/51 objects, amounting to 57 individual spectra once accounting for extractions from multiple \hii\ regions within the same system. Using SDSS $ugriz$ images, a photometric analysis was undertaken on the complete sample of 51 galaxies. The main results from this study are as follows:
\begin{enumerate}
\item Direct method oxygen abundances were obtained via electron temperature measurements for 29/51 objects and were found to range from 7.43$<$12+log(O/H)$<$8.01. Eleven of these 29 objects have O/H values below 0.1~\Zsol\, and are therefore classified as being extremely metal poor.

\item Using the same `direct method', nitrogen-to-oxygen ratios were calculated and found to lie between $-1.88<\log$(N/O)$<-1.39$, which are typical values for the metallicity range described above.

\item Total luminosities were found to be in the range 
$39 <\log(L_{\rm H\alpha}/{\rm erg~s^{-1}) }<41$, 
which translate to current star-formation rates between 1.1$\times10^{-2}$ and 33.4$\times10^{-2}$~\Msol/yr and a median of $3.5\times10^{-2}$~\Msol/yr. In consideration of their faintness, SFRs were calculated using the low-luminosity regime recalibration of the K98 relation by  \citet{Lee:2009}.

\item Absolute $B$-band luminosities are between $\sim$ $-10$ and $-18$~mag and $V$-band surface brightnesses are between 23 and 25~mag~arcsec$^{-1}$.

\item Using SED modelling, stellar masses were found to range from $\log(M_*/\Msol)\sim5$ to 9, with an average of $\log(M_*/\Msol)\sim7$.  Specific SFRs were found to lie in the range $-10.0<\log($SSFR/yr$^{-1}$)$<-6.5$, with a mean of $\log{\rm (SSFR)/yr}^{-1}\sim -9$, which fit within the star-forming sequence for low-mass galaxies. However, without deep optical and infrared imaging, such mass estimates are thought to have significant uncertainties.
\end{enumerate}

From the photometric and spectroscopic analysis of our significantly larger sample of BDD galaxies, we confirm the findings of paper I in that BDDs appear to be a population of dwarf galaxies that lie in between quiescent dIrrs and active starbursting BCD galaxies. By assessing the location of the BDD sample within the mass-metallicity relation we find that, unlike bright emission line XMP galaxies, BDDs do not appear to be experiencing an excessive amount of metal-poor gas accretion relative to their stellar mass. As such, the triggering mechanism behind their ongoing star-formation may instead be due to past interactions or previously ejected gas that has now cooled. Their diffuse and irregular morphology separates them from the compact BCD structure, or the typically cometary \textit{bursty}-XMP galaxies. With regards to this latter class, the XMP-BDD galaxies instead belong to a `diffuse yet active' XMP category. Overall, BDDs appear to be a sample of dwarf galaxies that bridge the dIrr+BCD galaxy populations with regards to physical and morphological properties - a population of actively star-forming galaxies where the \hii\ regions are not centrally clustered, i.e., dIrrs experiencing an \textit{active} stage of star-formation. In the broader scheme of dwarf galaxy classification, they fit within the class of low-surface brightness dIrr galaxies.

In order to fully assess the role that BDDs may play within dwarf galaxy evolution, three major pieces of the puzzle are still required. Firstly, deep optical and infrared imaging would enable accurate photometric profiling of their underlying host component and allow us to compare their light distributions with those of dIrrs and BCDs. Secondly, star-formation histories obtained from resolved stellar population studies of the closest BDDs, could provide significant insight into their past, and present mode of star-formation, alongside their chemical evolutionary history. Finally, kinematical information from IFU observations would enable us to explore how it is that star-formation can be actively ongoing in such random hap-hazard structures. In the future, we hope to extend our BDD sample by applying our search algorithms to data from southern hemisphere surveys, such as DES and VST-VISTA.

\section*{Acknowledgments}
The authors are sincerely grateful to Danielle Berg for discussions regarding chemical abundance calculations, Janice Lee for advice on calculating star-formation rates in the low-luminosity regime, and Matt Auger for assistance with SED modelling. We are grateful to the University of Arizona Observatory time assignment committee who awarded time to this programme, and thank the MMT telescope operators and staff for technical support.   BLJ acknowledges support from the Science \&\ Technology Facilities Council (STFC). The research leading to these results has received funding from the European Research Council under the European Union's Seventh Framework Programme (FP/2007-2013)/ERC Grant Agreement no. 308024. SK acknowledges financial support from the ERC.  DPS acknowledges support from the National Science Foundation  through the grant AST-1410155. EWO is partially supported by NSF grant AST1313006.   This research has made use of the NASA/IPAC Extragalactic Database (NED) which is operated by the Jet Propulsion Laboratory, California Institute of Technology, under contract with the National Aeronautics and Space Administration.
\bibliographystyle{mn2e}
\bibliography{references}
\clearpage

\appendix
\onecolumn
 \setcounter{table}{0}
\renewcommand{\thetable}{A.\arabic{table}}
 
\begin{landscape}
\section{Observed Fluxes and Line Intensities}

\begin{center}
\begin{table*}
\caption{Spectroscopic line fluxes and de-reddened line intensities (both relative to \hb$=100$) as measured from the spectra shown in Figure~\ref{fig:allspec}.}\label{tab:fluxes_all}

\begin{tabular}{|l|cc|cc|cc|cc|}
\hline
 & \multicolumn{2}{c|}{ JKB~1.0} & \multicolumn{2}{c|}{ JKB~1.1 }& \multicolumn{2}{c|}{ JKB~2.0} & \multicolumn{2}{c|}{ JKB~2.1} \\
\hline\hline
\foii\       3727.10 & 
     170$\pm$     27 &      199$\pm$     33 & 
    69.9$\pm$   10.2 &     69.9$\pm$   10.2 & 
    65.3$\pm$    9.6 &     77.1$\pm$   11.4 & 
    18.6$\pm$    3.0 &     24.4$\pm$    3.9 \\ 
\foii\       3729.86 & 
     140$\pm$     23 &      163$\pm$     27 & 
    57.3$\pm$    8.3 &     57.3$\pm$    8.3 & 
    54.2$\pm$    8.0 &     64.0$\pm$    9.6 & 
    17.6$\pm$    2.8 &     23.1$\pm$    3.8 \\ 
\hei\        3820.84 & 
--- & --- & 
--- & --- & 
--- & --- & 
--- & --- \\ 
\fneiii\       3870.16 & 
    41.8$\pm$    8.5 &     47.8$\pm$    9.7 & 
    33.5$\pm$    4.8 &     33.5$\pm$    4.8 & 
    6.11$\pm$   1.36 &     7.04$\pm$   1.57 & 
    15.9$\pm$    2.4 &     20.1$\pm$    3.0 \\ 
 \hei+H8       3889.75 & 
--- & --- & 
    19.1$\pm$    2.8 &     19.1$\pm$    2.8 & 
    14.3$\pm$    2.3 &     16.4$\pm$    2.7 & 
    13.7$\pm$    2.1 &     17.2$\pm$    2.7 \\ 
\heii\       3969.55 & 
--- & --- & 
    24.7$\pm$    3.6 &     24.7$\pm$    3.6 & 
    10.6$\pm$    1.9 &     12.0$\pm$    2.2 & 
    16.4$\pm$    2.4 &     20.2$\pm$    3.1 \\ 
\hei\        4027.49 & 
--- & --- & 
    2.31$\pm$   0.74 &     2.31$\pm$   0.74 & 
--- & --- & 
--- & --- \\
\hline
\hline
\end{tabular}
\end{table*}
\begin{table*}
\ContinuedFloat
\caption{-- continued.}
\begin{center}
\begin{footnotesize}
\begin{tabular}{|l|cc|cc|cc|cc|}
\hline
 & \multicolumn{2}{c|}{ JKB~3.0} & \multicolumn{2}{c|}{ JKB~3.1 }& \multicolumn{2}{c|}{ JKB~4.0} & \multicolumn{2}{c|}{ JKB~5.0} \\
\hline\hline
\foii\       3727.10 & 
    58.3$\pm$   12.9 &     58.3$\pm$   12.9 & 
     212$\pm$     32 &      308$\pm$     51 & 
    99.6$\pm$   14.3 &    126.7$\pm$   18.5 & 
    91.3$\pm$   13.8 &    110.1$\pm$   16.9 \\ 
\foii\       3729.86 & 
     101$\pm$     22 &      101$\pm$     22 & 
     193$\pm$     29 &      280$\pm$     47 & 
    71.4$\pm$   10.3 &     90.9$\pm$   13.4 & 
     105$\pm$     15 &      127$\pm$     19 \\ 
\hei\        3820.84 & 
--- & --- & 
--- & --- & 
--- & --- & 
    12.1$\pm$    2.6 &     14.3$\pm$    3.1 \\ 
\fneiii\       3870.16 & 
    33.1$\pm$    7.3 &     33.1$\pm$    7.3 & 
--- & --- & 
    11.8$\pm$    1.9 &     14.5$\pm$    2.3 & 
    8.41$\pm$   1.91 &     9.88$\pm$   2.25 \\ 
 \hei+H8       3889.75 & 
    18.1$\pm$    4.0 &     18.1$\pm$    4.0 & 
--- & --- & 
    9.86$\pm$   1.67 &    12.10$\pm$   2.07 & 
    7.61$\pm$   2.29 &     8.92$\pm$   2.70 \\ 
\heii\       3969.55 & 
    24.5$\pm$    5.4 &     24.5$\pm$    5.4 & 
--- & --- & 
    12.8$\pm$    1.9 &     15.5$\pm$    2.4 & 
    7.41$\pm$   1.94 &     8.57$\pm$   2.25 \\ 
\hei\        4027.49 & 
    1.51$\pm$   0.42 &     1.51$\pm$   0.42 & 
--- & --- & 
--- & --- & 
--- & --- \\
\hline
\hline
\end{tabular}
\label{tab:fluxes_all}
\end{footnotesize}
\end{center}

\end{table*}
\begin{table*}
\ContinuedFloat
\caption{-- continued.}
\begin{center}
\begin{footnotesize}
\begin{tabular}{|l|cc|cc|cc|cc|}
\hline
 & \multicolumn{2}{c|}{ JKB~7.0} & \multicolumn{2}{c|}{ JKB~8.11 }& \multicolumn{2}{c|}{ JKB~8.12} & \multicolumn{2}{c|}{ JKB~8.2} \\
\hline\hline
\foii\       3727.10 & 
     136$\pm$     19 &      148$\pm$     21 & 
    91.0$\pm$   13.2 &     96.3$\pm$   14.0 & 
     126$\pm$     34 &      126$\pm$     34 & 
    71.3$\pm$   15.3 &     71.3$\pm$   15.3 \\ 
\foii\       3729.86 & 
    78.1$\pm$   11.3 &     85.2$\pm$   12.4 & 
     135$\pm$     19 &      143$\pm$     20 & 
     376$\pm$     70 &      376$\pm$     70 & 
    84.2$\pm$   16.9 &     84.2$\pm$   17.0 \\ 
\hei\        3820.84 & 
--- & --- & 
--- & --- & 
--- & --- & 
--- & --- \\ 
\fneiii\       3870.16 & 
    26.7$\pm$    3.9 &     28.8$\pm$    4.2 & 
    29.7$\pm$    4.5 &     31.1$\pm$    4.7 & 
--- & --- & 
--- & --- \\ 
 \hei+H8       3889.75 & 
    14.5$\pm$    2.2 &     15.6$\pm$    2.4 & 
    12.3$\pm$    2.3 &     12.9$\pm$    2.4 & 
--- & --- & 
--- & --- \\ 
\heii\       3969.55 & 
    19.5$\pm$    2.9 &     20.9$\pm$    3.1 & 
    16.9$\pm$    2.8 &     17.6$\pm$    2.9 & 
--- & --- & 
--- & --- \\ 
\hei\        4027.49 & 
--- & --- & 
--- & --- & 
--- & --- & 
--- & --- \\
\hline
\hline
\end{tabular}
\label{tab:fluxes_all}
\end{footnotesize}
\end{center}

\end{table*}
\begin{table*}
\ContinuedFloat
\caption{-- continued.}
\begin{center}
\begin{footnotesize}
\begin{tabular}{|l|cc|cc|cc|cc|}
\hline
 & \multicolumn{2}{c|}{ JKB~8.3} & \multicolumn{2}{c|}{ JKB~15.1 }& \multicolumn{2}{c|}{ JKB~15.2} & \multicolumn{2}{c|}{ JKB~16.1} \\
\hline\hline
\foii\       3727.10 & 
    55.1$\pm$    8.1 &     55.1$\pm$    8.3 & 
    52.3$\pm$    7.8 &     57.4$\pm$    8.6 & 
     100$\pm$     14 &      120$\pm$     18 & 
    57.6$\pm$    8.4 &     57.6$\pm$    8.4 \\ 
\foii\       3729.86 & 
     100$\pm$     14 &      100$\pm$     14 & 
    57.2$\pm$    8.5 &     62.9$\pm$    9.3 & 
     101$\pm$     15 &      121$\pm$     18 & 
    98.9$\pm$   14.2 &     98.9$\pm$   14.2 \\ 
\hei\        3820.84 & 
--- & --- & 
--- & --- & 
--- & --- & 
--- & --- \\ 
\fneiii\       3870.16 & 
    25.3$\pm$    3.8 &     25.3$\pm$    3.9 & 
    27.8$\pm$    4.1 &     30.1$\pm$    4.4 & 
    20.6$\pm$    3.5 &     23.9$\pm$    4.0 & 
    25.1$\pm$    3.7 &     25.1$\pm$    3.7 \\ 
 \hei+H8       3889.75 & 
    13.7$\pm$    2.4 &     13.7$\pm$    2.4 & 
    15.0$\pm$    2.4 &     16.2$\pm$    2.6 & 
    13.0$\pm$    2.7 &     15.1$\pm$    3.2 & 
    18.6$\pm$    2.8 &     18.6$\pm$    2.8 \\ 
\heii\       3969.55 & 
    18.7$\pm$    3.0 &     18.7$\pm$    3.0 & 
    23.1$\pm$    3.4 &     24.8$\pm$    3.7 & 
    10.4$\pm$    2.3 &     11.9$\pm$    2.6 & 
    19.4$\pm$    2.9 &     19.4$\pm$    2.9 \\ 
\hei\        4027.49 & 
--- & --- & 
--- & --- & 
--- & --- & 
--- & --- \\
\hline
\hline
\end{tabular}
\label{tab:fluxes_all}
\end{footnotesize}
\end{center}

\end{table*}
\begin{table*}
\ContinuedFloat
\caption{-- continued.}
\begin{center}
\begin{footnotesize}
\begin{tabular}{|l|cc|cc|cc|cc|}
\hline
 & \multicolumn{2}{c|}{ JKB~16.2} & \multicolumn{2}{c|}{ JKB~16.3 }& \multicolumn{2}{c|}{ JKB~18.0} & \multicolumn{2}{c|}{ JKB~18.1} \\
\hline\hline
\foii\       3727.10 & 
    60.1$\pm$   20.2 &     69.3$\pm$   23.4 & 
     120$\pm$     28 &      137$\pm$     32 & 
     110$\pm$     16 &      128$\pm$     19 & 
     500$\pm$     91 &      788$\pm$    157 \\ 
\foii\       3729.86 & 
     281$\pm$     48 &      325$\pm$     56 & 
     266$\pm$     48 &      304$\pm$     56 & 
    86.6$\pm$   12.9 &    101.3$\pm$   15.2 & 
     218$\pm$     45 &      344$\pm$     76 \\ 
\hei\        3820.84 & 
    43.8$\pm$   12.8 &     49.9$\pm$   14.6 & 
--- & --- & 
    5.78$\pm$   1.71 &     6.66$\pm$   1.98 & 
     136$\pm$     27 &      205$\pm$     43 \\ 
\fneiii\       3870.16 & 
    31.9$\pm$   10.4 &     36.0$\pm$   11.8 & 
    33.1$\pm$   12.1 &     37.1$\pm$   13.6 & 
    23.3$\pm$    3.6 &     26.6$\pm$    4.1 & 
    47.3$\pm$   12.6 &     69.9$\pm$   19.2 \\ 
 \hei+H8       3889.75 & 
--- & --- & 
--- & --- & 
    15.9$\pm$    2.9 &     18.1$\pm$    3.3 & 
--- & --- \\ 
\heii\       3969.55 & 
--- & --- & 
--- & --- & 
    12.0$\pm$    2.2 &     13.5$\pm$    2.5 & 
--- & --- \\ 
\hei\        4027.49 & 
--- & --- & 
--- & --- & 
--- & --- & 
--- & --- \\
\hline
\hline
\end{tabular}
\label{tab:fluxes_all}
\end{footnotesize}
\end{center}

\end{table*}
\begin{table*}
\ContinuedFloat
\caption{-- continued.}
\begin{center}
\begin{footnotesize}
\begin{tabular}{|l|cc|cc|cc|cc|}
\hline
 & \multicolumn{2}{c|}{ JKB~21.0} & \multicolumn{2}{c|}{ JKB~22.0 }& \multicolumn{2}{c|}{ JKB~22.1} & \multicolumn{2}{c|}{ JKB~24.0} \\
\hline\hline
\foii\       3727.10 & 
     136$\pm$     22 &      160$\pm$     26 & 
    66.4$\pm$    9.5 &     76.1$\pm$   11.0 & 
    64.2$\pm$    9.3 &     73.6$\pm$   10.7 & 
     178$\pm$     25 &      276$\pm$     42 \\ 
\foii\       3729.86 & 
     197$\pm$     30 &      231$\pm$     36 & 
    71.6$\pm$   10.3 &     82.0$\pm$   11.8 & 
    80.2$\pm$   11.5 &     91.9$\pm$   13.3 & 
    73.4$\pm$   10.9 &    113.6$\pm$   17.8 \\ 
\hei\        3820.84 & 
--- & --- & 
--- & --- & 
--- & --- & 
    6.57$\pm$   1.76 &     9.75$\pm$   2.66 \\ 
\fneiii\       3870.16 & 
    40.0$\pm$    7.8 &     45.9$\pm$    9.0 & 
    32.3$\pm$    4.6 &     36.3$\pm$    5.2 & 
    31.1$\pm$    4.5 &     35.0$\pm$    5.1 & 
    30.7$\pm$    4.6 &     44.7$\pm$    6.9 \\ 
 \hei+H8       3889.75 & 
--- & --- & 
    13.9$\pm$    2.1 &     15.6$\pm$    2.3 & 
    13.4$\pm$    2.0 &     15.0$\pm$    2.3 & 
    8.42$\pm$   1.87 &    12.21$\pm$   2.76 \\ 
\heii\       3969.55 & 
    22.4$\pm$    5.7 &     25.4$\pm$    6.5 & 
    21.2$\pm$    3.1 &     23.6$\pm$    3.4 & 
    20.5$\pm$    3.0 &     22.8$\pm$    3.4 & 
--- & --- \\ 
\hei\        4027.49 & 
--- & --- & 
    2.15$\pm$   0.62 &     2.37$\pm$   0.68 & 
--- & --- & 
--- & --- \\
\hline
\hline
\end{tabular}
\label{tab:fluxes_all}
\end{footnotesize}
\end{center}

\end{table*}
\begin{table*}
\ContinuedFloat
\caption{-- continued.}
\begin{center}
\begin{footnotesize}
\begin{tabular}{|l|cc|cc|cc|cc|}
\hline
 & \multicolumn{2}{c|}{ JKB~24.1} & \multicolumn{2}{c|}{ JKB~25.0 }& \multicolumn{2}{c|}{ JKB~26.0} & \multicolumn{2}{c|}{ JKB~26.1} \\
\hline\hline
\foii\       3727.10 & 
     156$\pm$     22 &      197$\pm$     28 & 
    75.3$\pm$   10.7 &     75.3$\pm$   11.0 & 
    80.8$\pm$   12.0 &     81.5$\pm$   12.1 & 
    85.6$\pm$   12.3 &     90.7$\pm$   13.1 \\ 
\foii\       3729.86 & 
    61.4$\pm$    9.0 &     77.5$\pm$   11.6 & 
    74.8$\pm$   10.6 &     74.8$\pm$   11.0 & 
    50.8$\pm$    8.0 &     51.3$\pm$    8.0 & 
    61.9$\pm$    9.0 &     65.6$\pm$    9.6 \\ 
\hei\        3820.84 & 
    4.50$\pm$   1.17 &     5.56$\pm$   1.45 & 
--- & --- & 
--- & --- & 
--- & --- \\ 
\fneiii\       3870.16 & 
    25.6$\pm$    3.8 &     31.2$\pm$    4.6 & 
    26.6$\pm$    3.8 &     26.6$\pm$    3.9 & 
    30.4$\pm$    4.6 &     30.6$\pm$    4.7 & 
    18.9$\pm$    2.8 &     19.8$\pm$    2.9 \\ 
 \hei+H8       3889.75 & 
    18.2$\pm$    2.8 &     22.1$\pm$    3.4 & 
    14.0$\pm$    2.0 &     14.0$\pm$    2.1 & 
    9.74$\pm$   2.22 &     9.81$\pm$   2.24 & 
    14.9$\pm$    2.3 &     15.7$\pm$    2.4 \\ 
\heii\       3969.55 & 
    18.6$\pm$    2.9 &     22.3$\pm$    3.4 & 
    21.7$\pm$    3.1 &     21.7$\pm$    3.2 & 
--- & --- & 
    14.5$\pm$    2.2 &     15.1$\pm$    2.3 \\ 
\hei\        4027.49 & 
--- & --- & 
    1.44$\pm$   0.38 &     1.44$\pm$   0.38 & 
--- & --- & 
--- & --- \\
\hline
\hline
\end{tabular}
\label{tab:fluxes_all}
\end{footnotesize}
\end{center}

\end{table*}
\begin{table*}
\ContinuedFloat
\caption{-- continued.}
\begin{center}
\begin{footnotesize}
\begin{tabular}{|l|cc|cc|cc|cc|}
\hline
 & \multicolumn{2}{c|}{ JKB~28.0} & \multicolumn{2}{c|}{ JKB~29.0 }& \multicolumn{2}{c|}{ JKB~29.2} & \multicolumn{2}{c|}{ JKB~37.0} \\
\hline\hline
\foii\       3727.10 & 
    57.4$\pm$    8.5 &     57.4$\pm$    8.6 & 
    87.8$\pm$   13.7 &     96.0$\pm$   15.0 & 
    61.7$\pm$    9.7 &     66.3$\pm$   10.4 & 
    97.1$\pm$   14.0 &    130.0$\pm$   19.2 \\ 
\foii\       3729.86 & 
    51.9$\pm$    7.8 &     51.9$\pm$    7.8 & 
    97.5$\pm$   15.0 &    106.6$\pm$   16.4 & 
     116$\pm$     17 &      125$\pm$     18 & 
    75.6$\pm$   11.0 &    101.3$\pm$   15.1 \\ 
\hei\        3820.84 & 
--- & --- & 
--- & --- & 
--- & --- & 
--- & --- \\ 
\fneiii\       3870.16 & 
    13.8$\pm$    2.2 &     13.8$\pm$    2.2 & 
    10.4$\pm$    2.5 &     11.3$\pm$    2.7 & 
    68.2$\pm$   10.2 &     72.6$\pm$   10.8 & 
    26.5$\pm$    3.9 &     34.0$\pm$    5.1 \\ 
 \hei+H8       3889.75 & 
    10.7$\pm$    1.9 &     10.7$\pm$    1.9 & 
    12.4$\pm$    2.8 &     13.4$\pm$    3.0 & 
    6.20$\pm$   2.10 &     6.60$\pm$   2.23 & 
    12.0$\pm$    2.1 &     15.4$\pm$    2.7 \\ 
\heii\       3969.55 & 
    9.02$\pm$   1.73 &     9.02$\pm$   1.73 & 
--- & --- & 
    10.1$\pm$    2.4 &     10.7$\pm$    2.5 & 
    14.4$\pm$    2.4 &     18.0$\pm$    3.1 \\ 
\hei\        4027.49 & 
--- & --- & 
--- & --- & 
--- & --- & 
--- & --- \\
\hline
\hline
\end{tabular}
\label{tab:fluxes_all}
\end{footnotesize}
\end{center}

\end{table*}
\begin{table*}
\ContinuedFloat
\caption{-- continued.}
\begin{center}
\begin{footnotesize}
\begin{tabular}{|l|cc|cc|cc|cc|}
\hline
 & \multicolumn{2}{c|}{ JKB~40.0} & \multicolumn{2}{c|}{ JKB~42.0 }& \multicolumn{2}{c|}{ JKB~53.0} & \multicolumn{2}{c|}{ JKB~62.0} \\
\hline\hline
\foii\       3727.10 & 
    71.7$\pm$   11.8 &     79.9$\pm$   13.2 & 
    31.8$\pm$    4.8 &     34.0$\pm$    5.1 & 
    41.8$\pm$    6.0 &     42.0$\pm$    6.0 & 
    63.5$\pm$   11.1 &     84.7$\pm$   15.1 \\ 
\foii\       3729.86 & 
    98.5$\pm$   15.3 &    109.8$\pm$   17.1 & 
    30.0$\pm$    4.5 &     32.1$\pm$    4.8 & 
    42.9$\pm$    6.1 &     43.0$\pm$    6.1 & 
    59.3$\pm$   10.5 &     79.1$\pm$   14.3 \\ 
\hei\        3820.84 & 
--- & --- & 
--- & --- & 
--- & --- & 
--- & --- \\ 
\fneiii\       3870.16 & 
    23.1$\pm$    5.1 &     25.4$\pm$    5.6 & 
    48.9$\pm$    7.0 &     51.9$\pm$    7.5 & 
    26.0$\pm$    3.7 &     26.1$\pm$    3.7 & 
    24.5$\pm$    4.7 &     31.4$\pm$    6.1 \\ 
 \hei+H8       3889.75 & 
    17.3$\pm$    4.8 &     19.0$\pm$    5.3 & 
    17.9$\pm$    2.7 &     18.9$\pm$    2.9 & 
    17.9$\pm$    2.6 &     18.0$\pm$    2.6 & 
--- & --- \\ 
\heii\       3969.55 & 
    13.9$\pm$    3.3 &     15.2$\pm$    3.6 & 
    29.9$\pm$    4.3 &     31.5$\pm$    4.6 & 
    20.0$\pm$    2.8 &     20.1$\pm$    2.9 & 
--- & --- \\ 
\hei\        4027.49 & 
--- & --- & 
    2.02$\pm$   0.72 &     2.12$\pm$   0.76 & 
    1.24$\pm$   0.27 &     1.25$\pm$   0.27 & 
--- & --- \\
\hline
\hline
\end{tabular}
\label{tab:fluxes_all}
\end{footnotesize}
\end{center}

\end{table*}
\begin{table*}
\ContinuedFloat
\caption{-- continued.}
\begin{center}
\begin{footnotesize}
\begin{tabular}{|l|cc|cc|cc|cc|}
\hline
 & \multicolumn{2}{c|}{ JKB~64.0} & \multicolumn{2}{c|}{ JKB~69.0 }& \multicolumn{2}{c|}{ JKB~78.0} & \multicolumn{2}{c|}{ JKB~83.0} \\
\hline\hline
\foii\       3727.10 & 
    76.2$\pm$   11.2 &    100.5$\pm$   15.1 & 
    61.5$\pm$    9.2 &     64.6$\pm$    9.6 & 
    44.8$\pm$    6.6 &     44.8$\pm$    6.6 & 
     113$\pm$     16 &      124$\pm$     18 \\ 
\foii\       3729.86 & 
    59.8$\pm$    8.9 &     78.8$\pm$   12.1 & 
    63.4$\pm$    9.4 &     66.6$\pm$    9.9 & 
    46.9$\pm$    6.9 &     46.9$\pm$    6.9 & 
     138$\pm$     20 &      152$\pm$     22 \\ 
\hei\        3820.84 & 
    12.4$\pm$    2.3 &     15.9$\pm$    3.0 & 
--- & --- & 
--- & --- & 
--- & --- \\ 
\fneiii\       3870.16 & 
    15.3$\pm$    2.5 &     19.5$\pm$    3.2 & 
    35.2$\pm$    5.2 &     36.6$\pm$    5.4 & 
    22.4$\pm$    3.3 &     22.4$\pm$    3.3 & 
--- & --- \\ 
 \hei+H8       3889.75 & 
    4.89$\pm$   1.52 &     6.18$\pm$   1.92 & 
    12.8$\pm$    2.4 &     13.3$\pm$    2.5 & 
    16.8$\pm$    2.5 &     16.8$\pm$    2.5 & 
    23.5$\pm$    4.2 &     25.5$\pm$    4.6 \\ 
\heii\       3969.55 & 
--- & --- & 
    20.4$\pm$    3.3 &     21.1$\pm$    3.4 & 
    20.7$\pm$    3.0 &     20.7$\pm$    3.0 & 
    21.2$\pm$    3.8 &     22.8$\pm$    4.0 \\ 
\hei\        4027.49 & 
--- & --- & 
--- & --- & 
    2.49$\pm$   0.70 &     2.49$\pm$   0.70 & 
--- & --- \\
\hline
\hline
\end{tabular}
\label{tab:fluxes_all}
\end{footnotesize}
\end{center}

\end{table*}
\begin{table*}
\ContinuedFloat
\caption{-- continued.}
\begin{center}
\begin{footnotesize}
\begin{tabular}{|l|cc|cc|cc|cc|}
\hline
 & \multicolumn{2}{c|}{ JKB~85.0} & \multicolumn{2}{c|}{ JKB~89.0 }& \multicolumn{2}{c|}{ JKB~91.0} & \multicolumn{2}{c|}{ JKB~97.0} \\
\hline\hline
\foii\       3727.10 & 
    77.7$\pm$   12.1 &     93.6$\pm$   14.7 & 
     106$\pm$     15 &      116$\pm$     17 & 
     129$\pm$     18 &      133$\pm$     19 & 
    92.9$\pm$   13.7 &     99.7$\pm$   14.7 \\ 
\foii\       3729.86 & 
     112$\pm$     16 &      135$\pm$     20 & 
     120$\pm$     17 &      132$\pm$     19 & 
     129$\pm$     18 &      133$\pm$     19 & 
    87.2$\pm$   12.9 &     93.5$\pm$   13.8 \\ 
\hei\        3820.84 & 
    21.6$\pm$    4.2 &     25.6$\pm$    5.1 & 
    8.57$\pm$   2.14 &     9.30$\pm$   2.32 & 
--- & --- & 
--- & --- \\ 
\fneiii\       3870.16 & 
    22.9$\pm$    4.1 &     26.9$\pm$    4.8 & 
    33.8$\pm$    5.2 &     36.6$\pm$    5.6 & 
    31.9$\pm$    4.6 &     32.8$\pm$    4.7 & 
    23.2$\pm$    3.9 &     24.6$\pm$    4.1 \\ 
 \hei+H8       3889.75 & 
    17.0$\pm$    3.6 &     19.9$\pm$    4.2 & 
    12.2$\pm$    2.6 &     13.2$\pm$    2.9 & 
    17.5$\pm$    2.6 &     18.0$\pm$    2.7 & 
    14.7$\pm$    2.7 &     15.6$\pm$    2.9 \\ 
\heii\       3969.55 & 
--- & --- & 
    6.62$\pm$   2.07 &     7.10$\pm$   2.22 & 
    19.8$\pm$    2.9 &     20.2$\pm$    2.9 & 
    15.5$\pm$    2.7 &     16.4$\pm$    2.8 \\ 
\hei\        4027.49 & 
--- & --- & 
--- & --- & 
    2.94$\pm$   0.78 &     3.00$\pm$   0.80 & 
--- & --- \\
\hline
\hline
\end{tabular}
\label{tab:fluxes_all}
\end{footnotesize}
\end{center}

\end{table*}
\begin{table*}
\ContinuedFloat
\caption{-- continued.}
\begin{center}
\begin{footnotesize}
\begin{tabular}{|l|cc|cc|cc|cc|}
\hline
 & \multicolumn{2}{c|}{ JKB~97.1} & \multicolumn{2}{c|}{ JKB~119.0 }& \multicolumn{2}{c|}{ JKB~122.0} & \multicolumn{2}{c|}{ JKB~131.0} \\
\hline\hline
\foii\       3727.10 & 
     174$\pm$     40 &      222$\pm$     52 & 
     477$\pm$    125 &      731$\pm$    203 & 
    49.3$\pm$    7.0 &     55.1$\pm$    7.9 & 
     143$\pm$     21 &      181$\pm$     27 \\ 
\foii\       3729.86 & 
     112$\pm$     31 &      143$\pm$     40 & 
$<$   59.1  & $<$   90.7  & 
    61.4$\pm$    8.7 &     68.5$\pm$    9.8 & 
     145$\pm$     21 &      183$\pm$     28 \\ 
\hei\        3820.84 & 
--- & --- & 
--- & --- & 
--- & --- & 
--- & --- \\ 
\fneiii\       3870.16 & 
--- & --- & 
--- & --- & 
    21.4$\pm$    3.0 &     23.5$\pm$    3.4 & 
    36.7$\pm$    7.2 &     45.0$\pm$    8.9 \\ 
 \hei+H8       3889.75 & 
--- & --- & 
--- & --- & 
    15.5$\pm$    2.2 &     17.0$\pm$    2.4 & 
--- & --- \\ 
\heii\       3969.55 & 
--- & --- & 
--- & --- & 
    16.4$\pm$    2.3 &     17.8$\pm$    2.6 & 
--- & --- \\ 
\hei\        4027.49 & 
--- & --- & 
--- & --- & 
    1.04$\pm$   0.29 &     1.12$\pm$   0.32 & 
--- & --- \\
\hline
\hline
\end{tabular}
\label{tab:fluxes_all}
\end{footnotesize}
\end{center}

\end{table*}
\begin{table*}
\ContinuedFloat
\caption{-- continued.}
\begin{center}
\begin{footnotesize}
\begin{tabular}{|l|cc|cc|cc|cc|}
\hline
 & \multicolumn{2}{c|}{ JKB~133.0} & \multicolumn{2}{c|}{ JKB~133.1 }& \multicolumn{2}{c|}{ JKB~136.0} & \multicolumn{2}{c|}{ JKB~137.0} \\
\hline\hline
\foii\       3727.10 & 
     156$\pm$     25 &      170$\pm$     27 & 
    83.4$\pm$   12.6 &     94.1$\pm$   14.2 & 
    91.0$\pm$   14.2 &     91.0$\pm$   14.2 & 
     111$\pm$     25 &      186$\pm$     53 \\ 
\foii\       3729.86 & 
     189$\pm$     29 &      206$\pm$     32 & 
     125$\pm$     18 &      141$\pm$     20 & 
    72.4$\pm$   11.7 &     72.4$\pm$   11.7 & 
     223$\pm$     40 &      373$\pm$     93 \\ 
\hei\        3820.84 & 
--- & --- & 
--- & --- & 
--- & --- & 
--- & --- \\ 
\fneiii\       3870.16 & 
    15.7$\pm$    4.6 &     16.8$\pm$    5.0 & 
    21.8$\pm$    3.4 &     24.1$\pm$    3.8 & 
    16.0$\pm$    3.9 &     16.0$\pm$    3.9 & 
--- & --- \\ 
 \hei+H8       3889.75 & 
--- & --- & 
    9.40$\pm$   2.02 &    10.41$\pm$   2.24 & 
--- & --- & 
--- & --- \\ 
\heii\       3969.55 & 
--- & --- & 
    12.6$\pm$    2.3 &     13.8$\pm$    2.5 & 
--- & --- & 
--- & --- \\ 
\hei\        4027.49 & 
--- & --- & 
--- & --- & 
--- & --- & 
--- & --- \\
\hline
\hline
\end{tabular}
\label{tab:fluxes_all}
\end{footnotesize}
\end{center}

\end{table*}
\begin{table*}
\ContinuedFloat
\caption{-- continued.}
\begin{center}
\begin{footnotesize}
\begin{tabular}{|l|cc|cc|cc|cc|}
\hline
 & \multicolumn{2}{c|}{ JKB~138.2} & \multicolumn{2}{c|}{ JKB~141.0 }& \multicolumn{2}{c|}{ JKB~142.0} & \multicolumn{2}{c|}{ JKB~144.0} \\
\hline\hline
\foii\       3727.10 & 
     325$\pm$     49 &      493$\pm$    119 & 
     108$\pm$     15 &      131$\pm$     19 & 
    58.8$\pm$    8.9 &     70.2$\pm$   10.7 & 
    79.9$\pm$   11.3 &     91.1$\pm$   13.0 \\ 
\foii\       3729.86 & 
     147$\pm$     24 &      223$\pm$     55 & 
     135$\pm$     19 &      163$\pm$     23 & 
    83.8$\pm$   12.3 &    100.1$\pm$   14.8 & 
    62.7$\pm$    8.9 &     71.5$\pm$   10.2 \\ 
\hei\        3820.84 & 
    33.1$\pm$    7.3 &     48.2$\pm$   13.4 & 
--- & --- & 
    8.64$\pm$   2.04 &    10.15$\pm$   2.40 & 
--- & --- \\ 
\fneiii\       3870.16 & 
    43.6$\pm$    8.2 &     62.3$\pm$   15.5 & 
    26.8$\pm$    4.0 &     31.6$\pm$    4.7 & 
    15.0$\pm$    2.6 &     17.5$\pm$    3.0 & 
    36.0$\pm$    5.1 &     40.3$\pm$    5.7 \\ 
 \hei+H8       3889.75 & 
--- & --- & 
    18.6$\pm$    2.9 &     21.9$\pm$    3.4 & 
    12.9$\pm$    2.5 &     15.0$\pm$    2.9 & 
    17.0$\pm$    2.4 &     19.0$\pm$    2.7 \\ 
\heii\       3969.55 & 
--- & --- & 
    15.7$\pm$    2.6 &     18.2$\pm$    3.0 & 
    4.06$\pm$   1.47 &     4.66$\pm$   1.69 & 
    23.9$\pm$    3.4 &     26.4$\pm$    3.8 \\ 
\hei\        4027.49 & 
--- & --- & 
--- & --- & 
--- & --- & 
    1.97$\pm$   0.38 &     2.17$\pm$   0.42 \\
\hline
\hline
\end{tabular}
\label{tab:fluxes_all}
\end{footnotesize}
\end{center}

\end{table*}
\begin{table*}
\ContinuedFloat
\caption{-- continued.}
\begin{center}
\begin{footnotesize}
\begin{tabular}{|l|cc|cc|}
\hline
 & \multicolumn{2}{c|}{ JKB~144.1}   \\
\hline\hline
\foii\       3727.10 & 
     188$\pm$     50 &      277$\pm$     78 \\ 
\foii\       3729.86 & 
     256$\pm$     62 &      378$\pm$     97 \\ 
\hei\        3820.84 & 
--- & --- \\ 
\fneiii\       3870.16 & 
     129$\pm$     32 &      181$\pm$     46 \\ 
 \hei+H8       3889.75 & 
--- & --- \\ 
\heii\       3969.55 & 
--- & --- \\ 
\hei\        4027.49 & 
--- & --- \\
\hline
\hline
\end{tabular}
\label{tab:fluxes_all}
\end{footnotesize}
\end{center}

\end{table*}
\end{center}
\end{landscape}

\onecolumn
 \setcounter{figure}{0}
\renewcommand{\thefigure}{B.\arabic{figure}}
\section{Observed spectra}
MMT spectra of the sample of BDD galaxies detailed in Table~\ref{tab:gals} and shown in Fig.~\ref{fig:gals}. Wavelengths are given in the rest frame of each galaxy. Each spectrum has been smoothed with a 5-pixel boxcar for presentation purposes. Dashed lines indicate emission lines detected in each spectrum, as listed in Table B1. Object titles listed with `$\star$' highlight those for which no strong emission lines were detected and are hence shown at the observed wavelength.
\begin{figure*}
\includegraphics[scale=0.9]{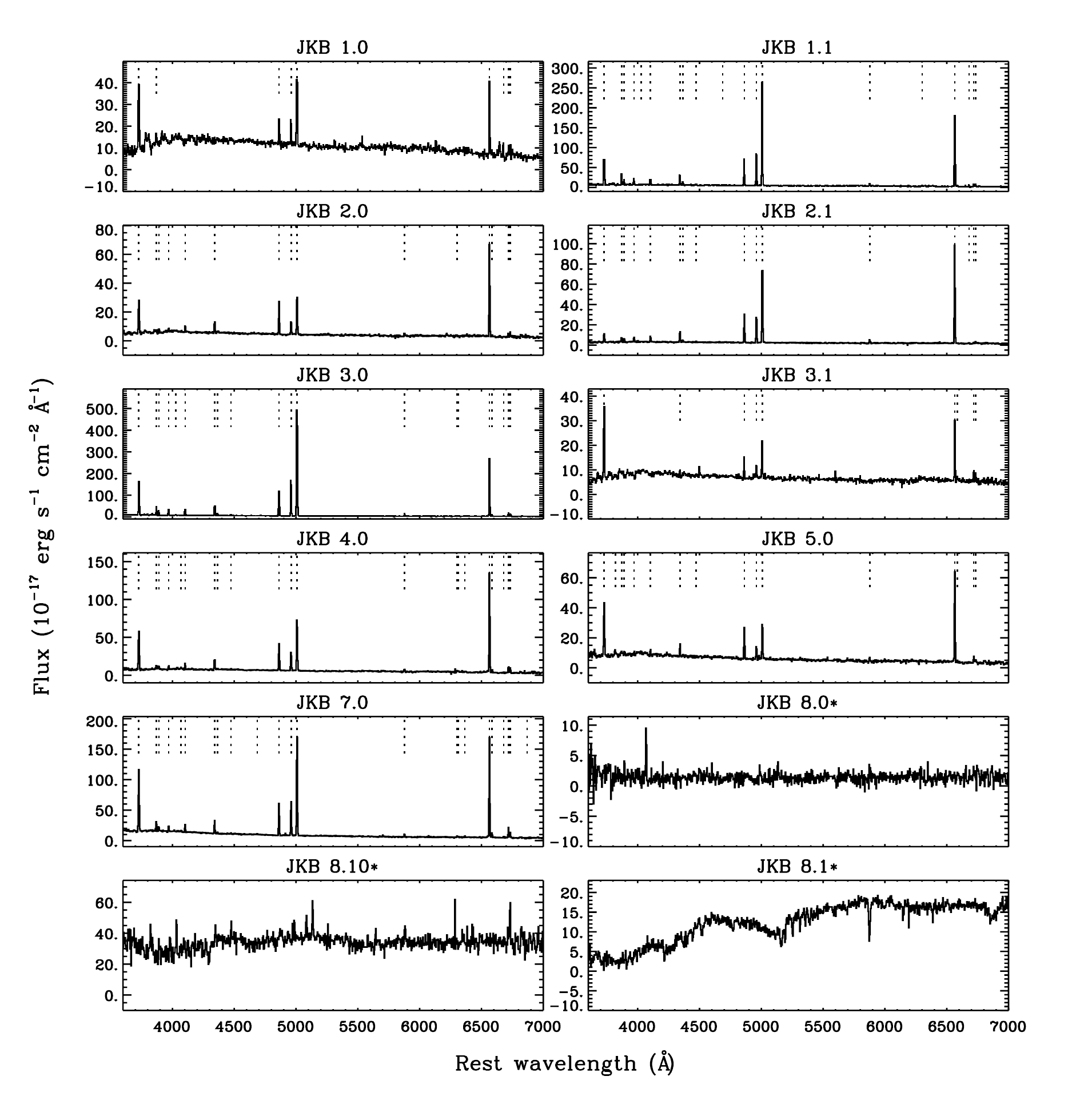}
\caption{}\label{fig:allspec}
\end{figure*} 
\begin{figure*}
\ContinuedFloat
\includegraphics[scale=0.9]{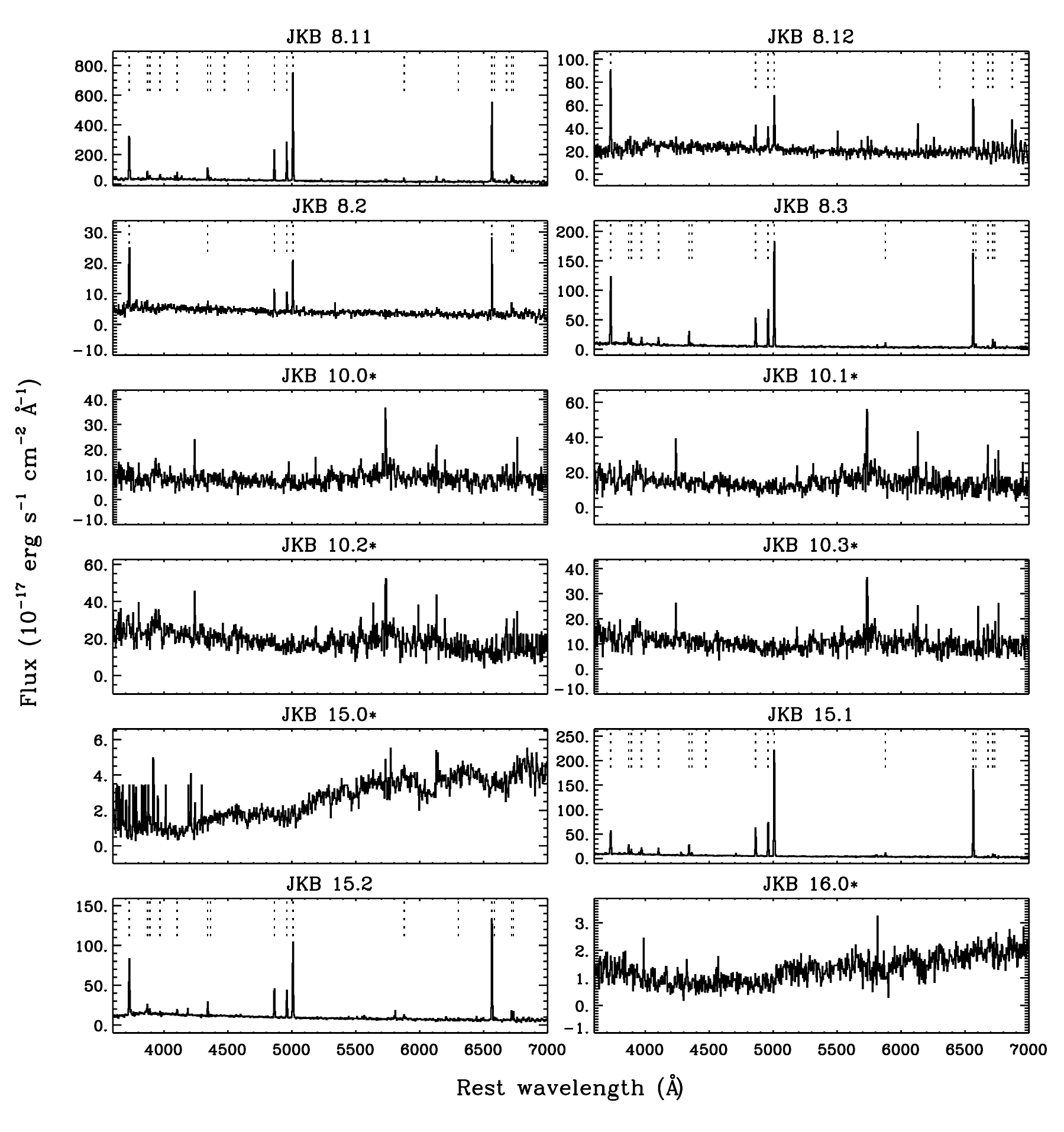}
\caption{-- continued.} 
\end{figure*} 
\begin{figure*}
\ContinuedFloat
\includegraphics[scale=0.9]{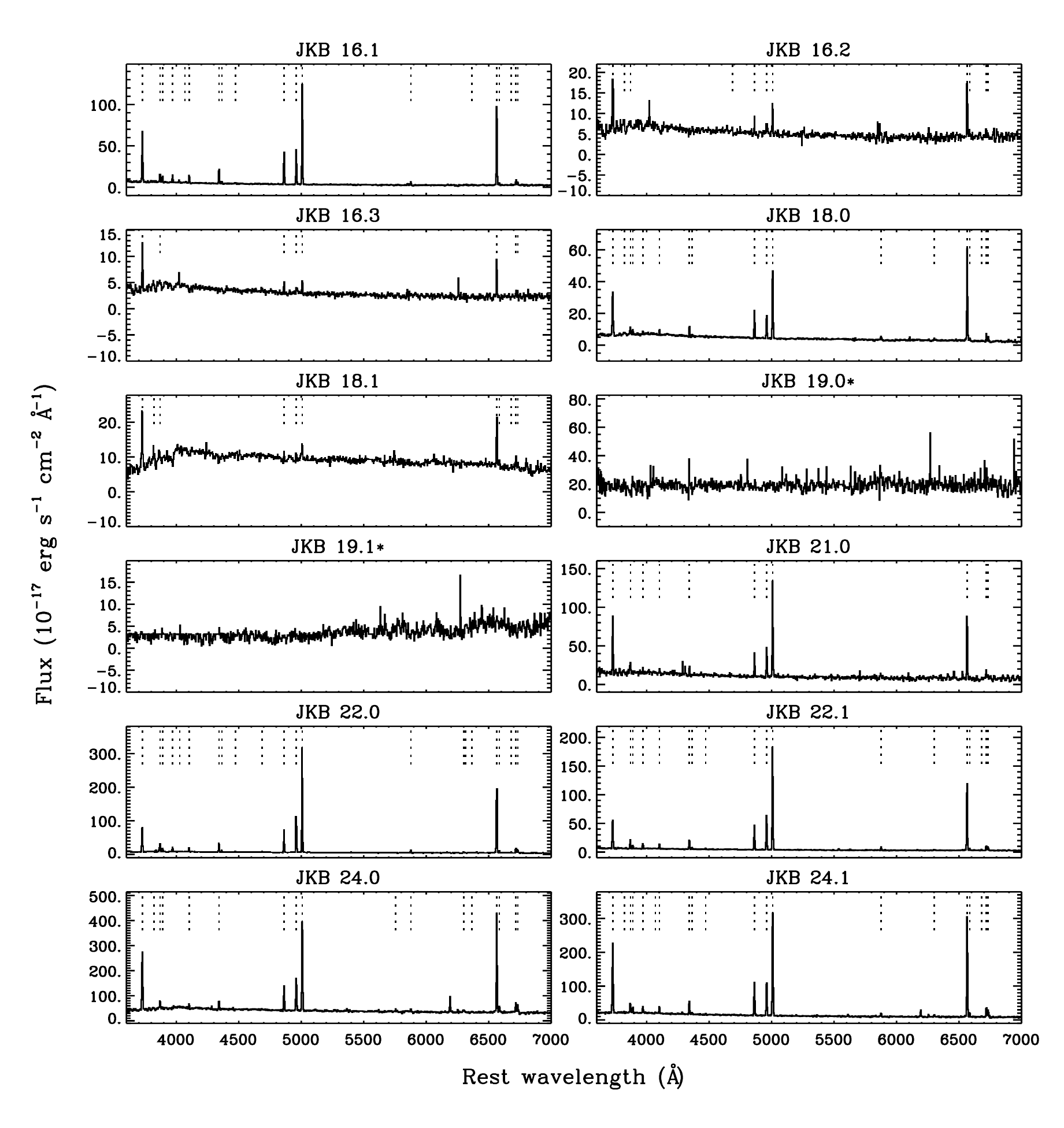}
\caption{ -- continued.} 
\end{figure*} 
\begin{figure*}
\ContinuedFloat
\includegraphics[scale=0.9]{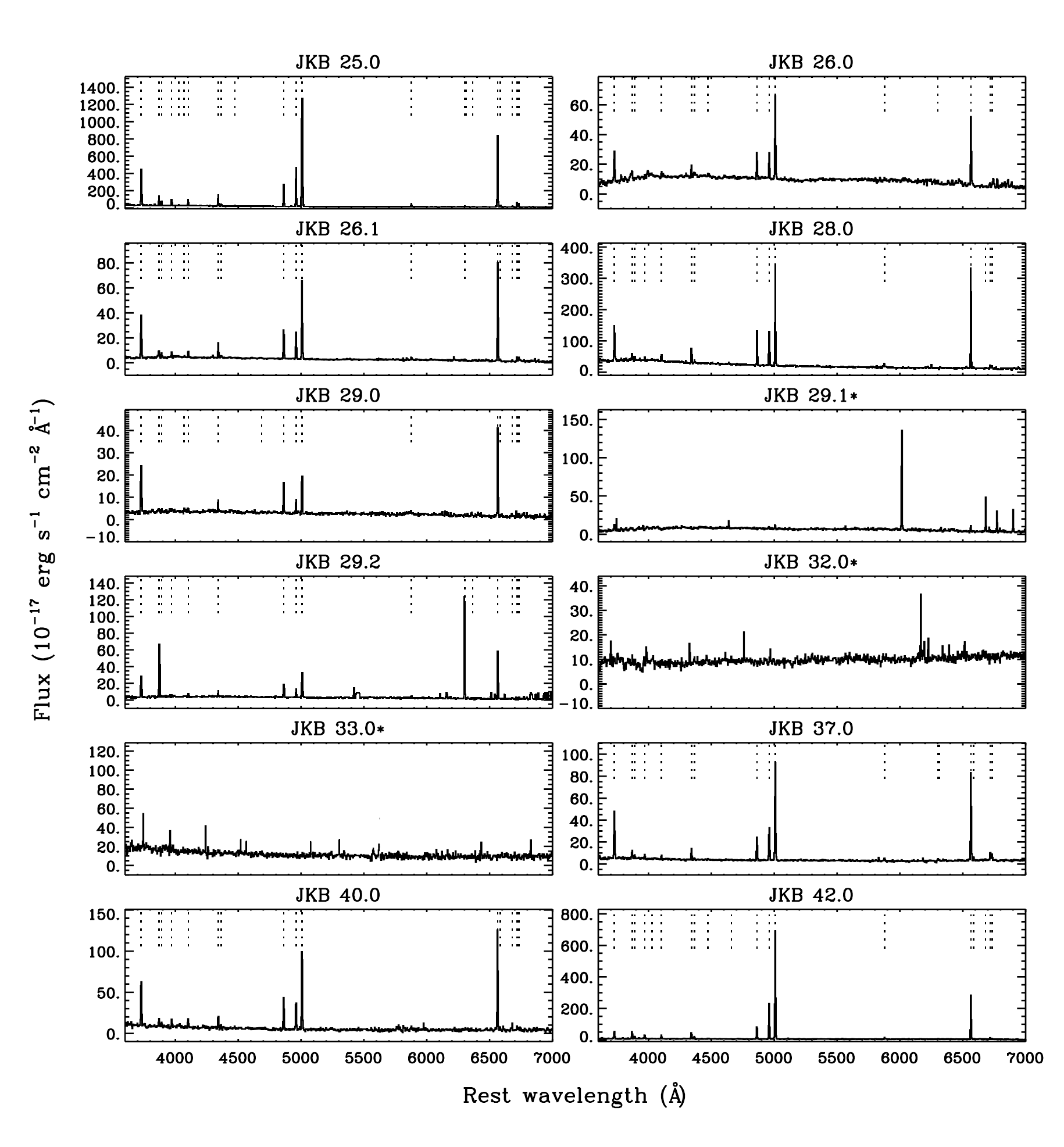}
\caption{ -- continued.} 
\end{figure*} 
\begin{figure*}
\ContinuedFloat
\includegraphics[scale=0.9]{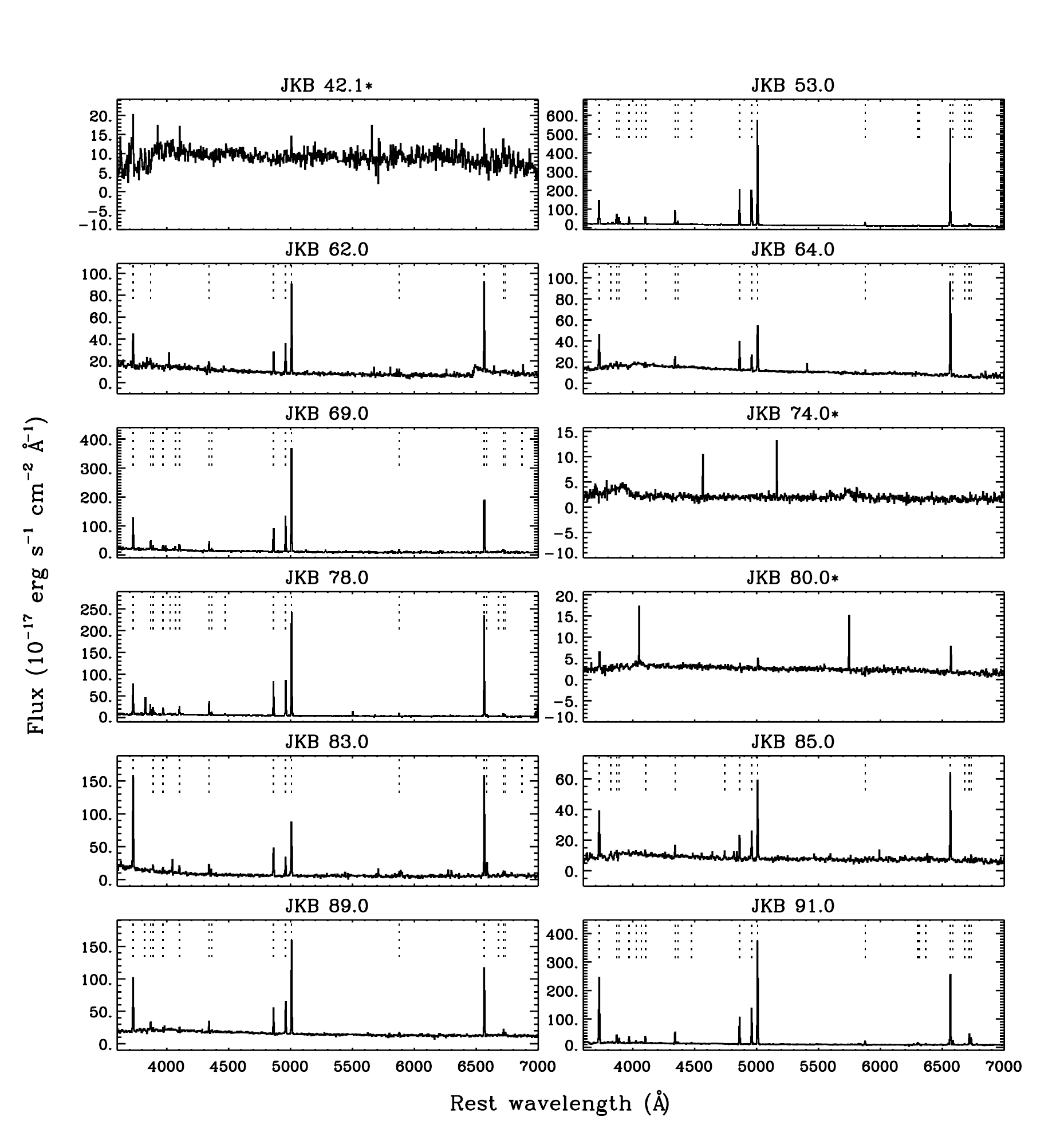}
\caption{-- continued. } 
\end{figure*} 
\begin{figure*}
\ContinuedFloat
\includegraphics[scale=0.9]{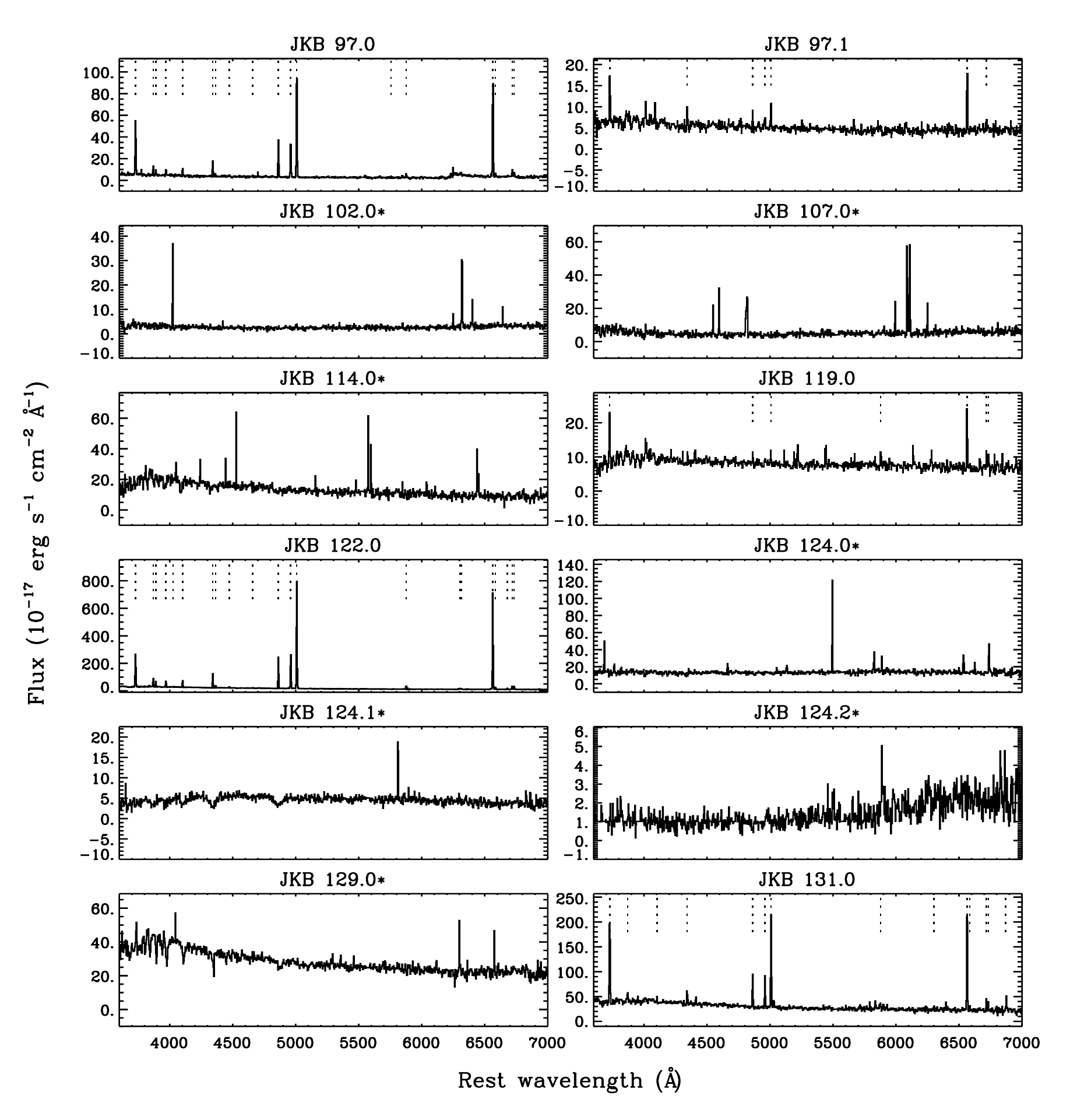}
\caption{ -- continued.} 
\end{figure*} 
\begin{figure*}
\ContinuedFloat
\includegraphics[scale=0.9]{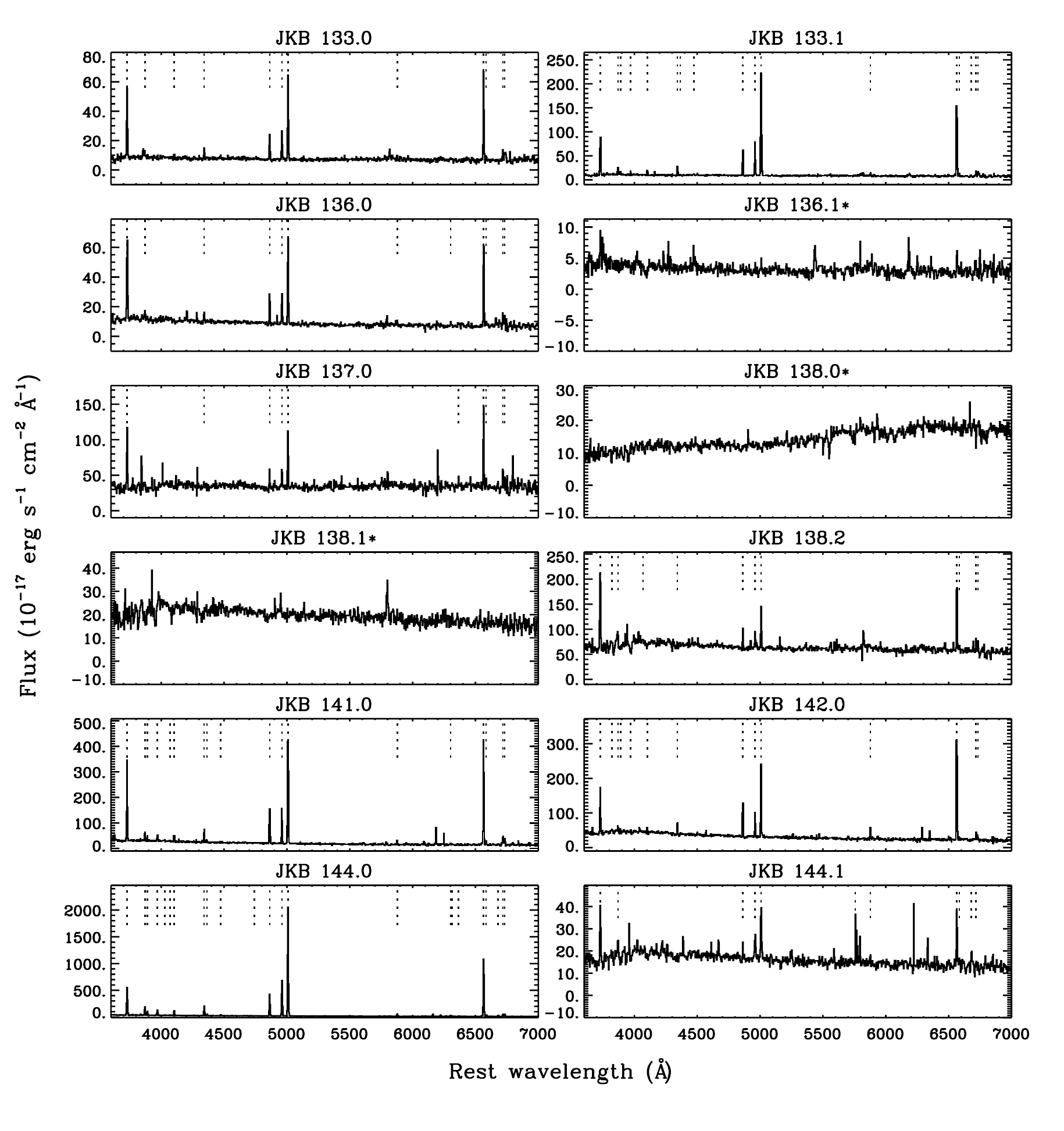}
\caption{ -- continued.} 
\end{figure*}

\bsp

\label{lastpage}

\end{document}